\documentclass[aps,prb,showpacs,twocolumn,reprint,superscriptaddress,longbibliography,showkeys]{revtex4-2}
\usepackage{amsmath}
\usepackage{amssymb}
\usepackage{graphicx}
\usepackage{color}
\usepackage{multirow}
\usepackage{dcolumn}
\usepackage{textcomp}
\usepackage{longtable}
\usepackage{makecell}
\usepackage{booktabs}
\usepackage{array}
\usepackage{titlesec}
\usepackage{hyperref}
\hypersetup{hypertex=true, colorlinks=true, linkcolor=blue, anchorcolor=blue,citecolor=blue,urlcolor=blue}
\usepackage{bm}
\PassOptionsToPackage{english}{babel}

\begin{document}

\title{Ferroelectrically switchable magnetic multistates in MnBi$_2$Te$_4$(Bi$_2$Te$_3$)$_n$ and MnSb$_2$Te$_4$(Sb$_2$Te$_3$)$_n$ ($n$ = 0, 1) thin films}
\author {Guoliang Yu}
\affiliation{Key Laboratory for Matter Microstructure and Function of Hunan Province,
Key Laboratory of Low-Dimensional Quantum Structures and Quantum Control of Ministry of Education, School of Physics and Electronics, Hunan Normal University, Changsha 410081, China}

\author {Chuhan Tang}
\affiliation{Key Laboratory for Matter Microstructure and Function of Hunan Province,
Key Laboratory of Low-Dimensional Quantum Structures and Quantum Control of Ministry of Education, School of Physics and Electronics, Hunan Normal University, Changsha 410081, China}

\author {Zhiqiang Tian}
\affiliation{Key Laboratory for Matter Microstructure and Function of Hunan Province,
Key Laboratory of Low-Dimensional Quantum Structures and Quantum Control of Ministry of Education, School of Physics and Electronics, Hunan Normal University, Changsha 410081, China}

\author {Ziming Zhu}
\affiliation{Key Laboratory for Matter Microstructure and Function of Hunan Province,
Key Laboratory of Low-Dimensional Quantum Structures and Quantum Control of Ministry of Education, School of Physics and Electronics, Hunan Normal University, Changsha 410081, China}

\author {Peitao Liu}
\affiliation{Shenyang National Laboratory for Materials Science, Institute of Metal Research, Chinese Academy of Sciences, 110016 Shenyang, China}
\affiliation{School of Materials Science and Engineering, University of Science and Technology of China, 110016 Shenyang, China}
\date{\today}

\author {Anlian Pan}
\affiliation{Key Laboratory for Micro-Nano Physics and Technology of Hunan Province, College of Materials Science and Engineering, Hunan University, Changsha 410082, China}

\author {Mingxing Chen}
\email{mxchen@hunnu.edu.cn}
\affiliation{Key Laboratory for Matter Microstructure and Function of Hunan Province,
Key Laboratory of Low-Dimensional Quantum Structures and Quantum Control of Ministry of Education, School of Physics and Electronics, Hunan Normal University, Changsha 410081, China}

\author {Xing-Qiu Chen}
\affiliation{Shenyang National Laboratory for Materials Science, Institute of Metal Research, Chinese Academy of Sciences, 110016 Shenyang, China}
\affiliation{School of Materials Science and Engineering, University of Science and Technology of China, 110016 Shenyang, China}
\date{\today}

\begin{abstract}
Ferroelectric control of two-dimensional magnetism is promising in fabricating electronic devices with high speed and low energy consumption. The newly discovered layered MnBi$_2$Te$_4$(Bi$_2$Te$_3$)$_n$ and their Sb counterparts exhibit A-type antiferromagnetism with intriguing topological properties. Here, we propose to obtain tunable magnetic multistates in their thin films by ferroelectrically manipulating the interlayer magnetic couplings (IMCs) based on the Heisenberg model and first-principles calculations. Our strategy relies on that interfacing the thin films with appropriate ferroelectric materials can switch on/off an interlayer hopping channel between Mn-$e_g$ orbitals as the polarizations reversed, thus resulting in a switchable interlayer antiferromagnetism-to-ferromagnetism transition. On the other hand, the interface effect leads to asymmetric energy barrier heights for the two polarization states. These properties allow us to build ferroelectrically switchable triple and quadruple magnetic states with multiple Chern numbers in thin films. Our study reveals that ferroelectrically switchable magnetic and topological multistates in MnBi$_2$Te$_4$ family can be obtained by rational design for multifunctional electronic devices, which can also be applied to other two-dimensional magnetic materials.
\end{abstract}


\maketitle
\section{INTRODUCTION}
Two-dimensional (2D) magnetic materials provide an ideal platform to explore novel magnetic and electronic properties~\cite{Huang2017,Gong2017,Fei2018,Burch2018-ix, Gong2019-lo,Wang2021,Ni2021}. The delicate interlayer exchange couplings in these systems enable a variety of methods of manipulating their magnetism. For instance, recent studies revealed that the A-type antiferromagnetic (AFM) CrI$_3$ bilayer could be tuned into ferromagnetic (FM) by external electric field~\cite{huang2018electrical,xu2020electric}, electrostatic doping~\cite{jiang2018controlling,soriano2020magnetic}, and interface engineering~\cite{Lu2020,Cheng2021,yang2021realization,shen2021nonvolatile,Tian2021CrI3,li2022intriguing}. Twisting the bilayer may yield non-collinear magnetism~\cite{tong2018skyrmions,song2021direct,akram2021moire}.

Recently, the MnBi$_2$Te$_4$ family, i.e., MnBi$_2$Te$_4$(Bi$_2$Te$_3$)$_n$ and MnSb$_2$Te$_4$(Bi$_2$Te$_3$)$_n$, which are hereafter referred to as MAT, have attracted much attention due to the coexistence and interesting interplay of intrinsic magnetism and band topology in them~\cite{Otrokov2019,Zhang2019,Li2019,Gong2019,Hao2019surface,Zhang2019NM,deng2020quantum,yan2019crystal,Zhang2019NM,ge2020high,Hu2020,
Li2020,zhou2020topological,WuMnBi4Te72020,Lian2020twist, Yang2021,zhang2021tunable,zhang2022nonreciprocal, Jianghua2022}. This series of materials also have a layered van der Waals (vdW) structure with the A-type AFM structure as revealed by experiments~\cite{Vidal2019Atype,Zeugner2019MnBi2Te4,Sass2020Atype, Yan2020Atype}, which preserves the combination of the time-reversal and a half lattice translation symmetry. As a result, many of them show nontrivial topological properties such as topological insulators~\cite{Otrokov2019} and axion insulators~\cite{Zhang2019}. Moreover, the systems can also be turned into Weyl semimetals as the interlayer couplings become ferromagnetic~\cite{Zhang2019,Li2019,zhang2021tunable}.

The A-type AFM coupling in MAT yields unusual even-odd layer-number dependent magnetism for their thin films~\cite{Otrokov2019PRL,Yang2021,zang2022layer}. The even-number (even-$N$) systems are expected to have no net magnetization. Whereas those with odd-number (odd-$N$) layers have uncompensated magnetization. This difference can lead to distinct topological properties for them. For instance, the thin films of MnBi$_2$Te$_4$ with odd-$N$ septuple layers are quantum anomalous Hall insulators. However, those with even-$N$ layers have a zero Chern number\citep{Otrokov2019PRL}. Instead, their topological properties can be characterized by the so-called pseudospin Chern number\citep{spin_Chern}. Due to the weak interlayer interaction, small magnetic fields could induce spin-flip transitions, giving rise to an AFM-to-FM transition in the IMCs~\cite{Sass2020Atype, Yang2021}. Chemical dopings~\cite{Han2021doping,riberolles2021evolution} and antisite defects~\cite{murakami2019realization,liu2021site,Du_tuning_2021,Garnica_native_2022,Liu_visualizing_2022} can also be used to manipulate the IMCs in these systems, although they may complicate the nature of the surface states. First-principles calculations suggest that the AFM double-septuple MnBi$_2$Te$_4$ could be driven into a Chern insulator with a high Chern number under electric fields~\cite{Du2020}, and could be driven into FM IMCs by interfacing with monolayers of transition-metal dichalcogenide and h-BN~\citep{Gao2021}.

In this paper, we propose to ferroelectrically tune the IMCs in MnBi$_2$Te$_4$(Bi$_2$Te$_3$)$_n$ and MnSb$_2$Te$_4$(Sb$_2$Te$_3$)$_n$ ($n$ = 0 and 1) thin films for magnetic multistates by interface, which is desired for memory devices with high density storage, high speed, and low power consumption. We reveal that hole doping can lead to an interlayer AFM-to-FM transition in MAT bilayers based on the understanding of the IMCs using the Heisenberg model. We provide a guideline for designing ferroelectric substrates that may induce transitions in the interlayer exchange couplings, i.e., polarization dependent IMCs, as demonstrated by our first-principles calculations. Moreover, we find that the interface effect results in symmetry breaking in the two polarization states of the FE substrate. This asymmetry allows us to design switchable magnetic multistates in sandwich structures made of MnBi$_2$Te$_4$ multilayers and 2D ferroelectric (FE) materials, which also exhibit distinct electronic and topological properties.

\section{COMPUTATIONAL DETAILS}
We perform DFT calculations for our systems using the Vienna Ab initio Simulation Package~\cite{kresse1996}. The pseudopotentials were constructed by the projector augmented wave method~\cite{bloechl1994,kresse1999}. An 11 $\times$ 11 $\times$ 1 and 21 $\times$ 21 $\times$ 1 $\Gamma$-centered $k$-mesh were used to sample the 2D Brillouin zone for structural relaxation and electronic structure calculations, respectively. The plane-wave energy cutoff is set to 400 eV for all calculations. A 20 \AA~vacuum region was used between adjacent plates to avoid the interaction between neighboring periodic images. Van der Waals (vdW) dispersion forces between the adsorbate and the substrate were accounted for through the DFT-D3~\cite{DFT-D3}. Different vdW methods/functionals such as DFT-D2 and optPBE-vdW were also used for comparison~\cite{DFT-D2,klimes2010,klimes2011}. The systems were fully relaxed until the residual force on each atom is less than 0.01 eV/\AA. The DFT+U method~\cite{Dudarev1998} is used to treat electron correlations due to the partially filled $d$-orbital of Mn for which a value of 5.34 eV is used~\cite{Otrokov2019}. Our results on the structural properties, magnetism, and band structures of the free-standing MAT films are consistent with previous studies~\cite{Otrokov2019,zhang2021tunable,zhou2020topological}. The kinetic pathways of transitions between different polarization states are calculated using the climbing image nudge elastic band (CI-NEB) method~\cite{Henkelman2000,henkelman2000improved}. The topological properties calculations were done using the WANNIER90~\cite{wannier90} and WannierTools package~\cite{WU2017}.

\section{RESULTS AND DISCUSSIONS}
\subsection{Concept of ferroelectric tuning of IMCs in MAT bilayers}
\begin{figure*}[t]
  \includegraphics[width=.95\linewidth]{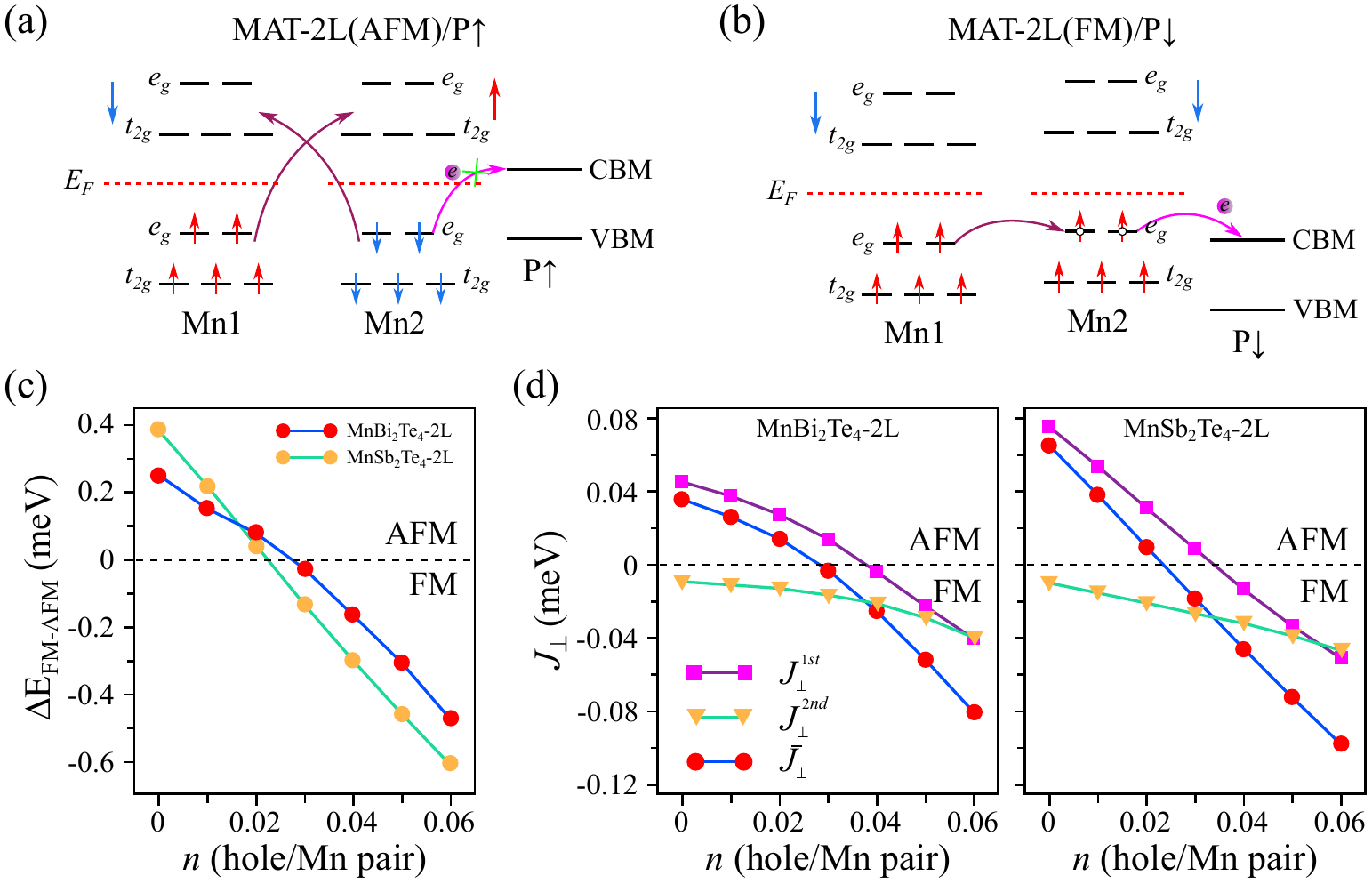}
  \caption{Interface engineering of the IMCs in MAT bilayers. (a) Spin states of Mn ions for the AFM interlayer coupling.  In the presence of a substrate that has a type-I or type-II band alignment with the bilayer, the IMCs remain AFM. (b) The IMCs becomes FM when there is a type-III band alignment between them so that the CBM of the substrate lower than the VBM of the MAT bilayer. In (b), the white circles denote hole dopings to the Mn-$e_g$ orbital. In (a, b), arrows denote spins. The dark and light red curves with an arrow denote hopping channels. The one marked by a cross means that electron hoppings are prohibited.  We use a FE material as the substrate, whose polarizations are labeled by $P$. $P\uparrow$ and $P\downarrow$ represent the up and down polarizations, respectively. (c) Energy difference between the FM and AFM states as a function of hole doping for freestanding MnBi$_2$Te$_4$ and MnSb$_2$Te$_4$ bilayers.  $\Delta E = E_{FM} - E_{AFM}$. (d) Doping dependence of $J_{\perp}$ for the two systems. $\bar{J_{\perp}} = J_{\perp}^{1st} + J_{\perp}^{2nd}$.}
 \label{fig1}
\end{figure*}

We begin by presenting the concept of FE tuning of IMCs in MAT bilayers, which is shown in Fig.~\ref{fig1}. In these systems, each Mn atom is coordinated with six chalcogen atoms, which form a distorted octahedron. The Mn-3$d$ orbitals are split into triply degenerate $t_{2g}$ states and the doubly degenerate $e_g$ states due to the octahedral ligand field. The states are further split due to
the magnetic exchange interaction between the Mn atoms. As a result, the majority states of the $t_{2g}$ and $e_g$ orbitals are fully occupied by the five $d$ electrons of
the Mn$^{2+}$ ions (see Fig.~\ref{fig1}), resulting in a high spin state for the Mn$^{2+}$ ions. Like the 2D magnetic bilayers reported by Refs \onlinecite{Li2020,Xiao_2020}, this type of occupation favors AFM IMCs between the Mn$^{+2}$ ions, which are mediated by the $p$-orbitals of the nonmetallic atoms (denoted by $\{p\dots p\}$). Whereas FM IMCs are energetically unfavorable because the $e_g^{\uparrow}-\{p\dots p\}-e_g^{\uparrow}$ hopping between the Mn-$d$ orbitals of adjacent layers is prohibited \cite{Li2020,Xiao_2020,fu_exchange_2020}. For our systems, reducing the occupation of the $d$ orbitals makes the $e_g^{\uparrow}-\{p\dots p\}-e_g^{\uparrow}$ hopping channel energetically favorable, thus enhancing the stability of the FM IMCs. Indeed,
our DFT calculations indicate that all the MAT bilayers undergo the AFM-to-FM transition by small hole dopings (see Fig.~\ref{fig1}c and Fig.~S1\cite{SM}), which is also expected for their multilayers.

The IMCs in MAT-2L can be understood using the following spin Hamiltonian.
\begin{equation}
H=\sum_{i,j} J_{\parallel}^{t}\mathbf{S_{i}} \cdot \mathbf{S_{j}}  + \sum_{m,n} J_{\parallel}^{b} \mathbf{S_{m}} \cdot \mathbf{S_{n} } + \sum_{i,m} J_{\perp} \mathbf{S_{i}} \cdot \mathbf{S_{m}},
\end{equation}
where $J_{\parallel}$ and $J_{\perp}$ denote the intra- and interlayer exchange interactions between the Mn ions, respectively. The intralayer ones are denoted by $J_{\parallel}^{t}$ for the top layer and $J_{\parallel}^{b}$ for the bottom layer, for which only the first nearest-neighbor interactions are taken into account. $J_{\parallel}^{t}$ are equal to $J_{\parallel}^{b}$ for the freestanding MAT-2L. Whereas for the interlayer ones, the second nearest neighbors are included. For the bilayers without doping, we obtain positive $J_{\perp}^{1st}$ and negative $J_{\perp}^{2nd}$ (see Fig.~\ref{fig1}d, Fig.~S2 and Table S1\cite{SM}).  Note that the magnitude of $J_{\perp}^{1st}$ is larger than that of $J_{\perp}^{2nd}$. So, the sum of $J_{\perp}^{1st}$ and $J_{\perp}^{2nd}$, i.e., $\bar{J_{\perp}} = J_{\perp}^{1st} + J_{\perp}^{2nd}$, is positive, which gives rise to AFM IMCs. Introducing hole doping suppresses $J_{\perp}^{1st}$ while enhances $J_{\perp}^{2nd}$. As a result, $\bar{J_{\perp}}$ decreases with increasing of the hole doping and eventually changes its sign across the AFM-to-FM transition (see Fig.~\ref{fig1}d).

Experiments find that most of MnA$_2$Te$_4$(A$_2$Te$_3$)$_n$ (A = Bi, Sb; $n$ = 0 $-$ 2) show n-type metallic conductivity mainly due to the existence of antisite mixing of A and Mn\cite{yan2019crystal, Zeugner2019MnBi2Te4, murakami2019realization, liu2021site,YanPRB2019, riberolles2021evolution}, which is unfavorable for the FM IMCs. Doping with Sb can induce a n-to-p transition in the carrier type\cite{murakami2019realization, YanPRB2019, riberolles2021evolution}. However, different groups report different IMCs at high concentrations of Sb, e.g.,  MnSb$_2$Te$_4$.  Both AFM and FM IMCs were reported by previous studies, which suggests that the IMCs in this family are strongly dependent on the growth condition\cite{murakami2019realization, YanPRB2019, riberolles2021evolution}.  In addition, recent experiment finds that doping of Sb at $x$ = 0.25 can give rise to an AFM-to-FM transition in the interlayer couplings for For Mn(Bi$_{1-x}$Sb$_x$)$_6$Te$_{10}$ \cite{riberolles2021evolution,Xie_2022}.
For the MAT thin films, the hole doping can
also be achieved via interfacial charge transfer which requires suitable band alignments between them and the substrates. When their bands are in the type-I or
type-II alignment, interfacial charge transfer can be negligible. In these cases, the IMCs are most likely to be AFM.
In contrast, electrons will be transferred from the MAT bilayer to the substrate when they are in the type-III band alignment that the valence band maximum (VBM) of the MAT bilayer are higher than the conduction band minimum (CBM) of the
substrate. Namely, hole doping is introduced to the MAT bilayer, which is desired for the AFM-to-FM transition. Ferroelectrically switchable IMCs may be achieved if a 2D FE materials serves as the substrate so that reversing its polarizations gives rise to a switching of the band alignment from type-III to type-I (II) or vice versa (see Fig.~\ref{fig1}). However, one can expect that the transferred electrons mainly come from the interfacial MAT layer because of the vdW-type interlayer bonding. Therefore, the spin-flipping most likely happen to the interfacial MAT layer rather than those further away from the substrate.

\subsection{MAT-2L/In$_2$Se$_3$ heterostructures}

\begin{figure}[t]
  \includegraphics[width=.95\linewidth]{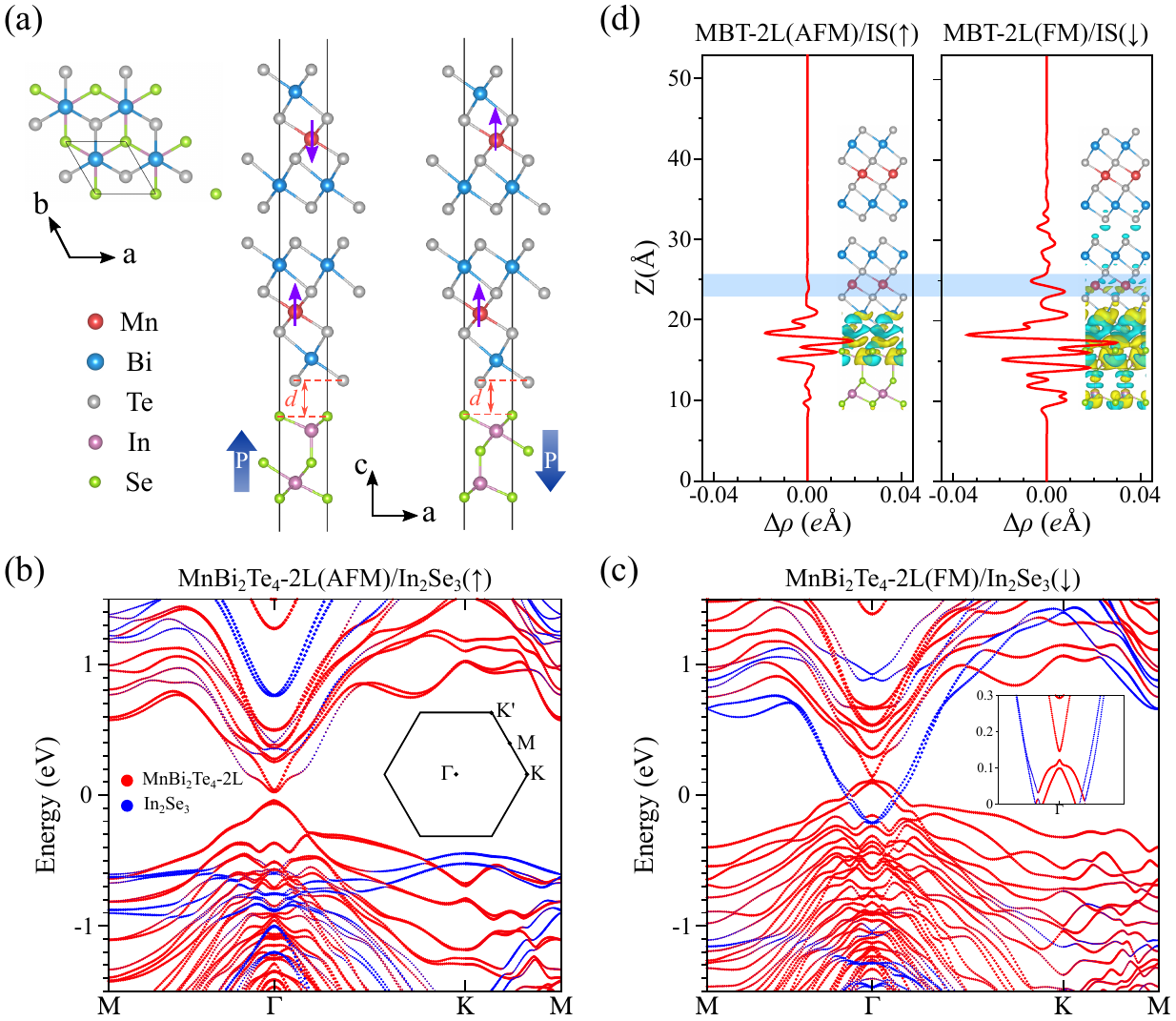}
  \caption{Ferroelectric control of AFM-to-FM transition in MnBi$_2$Te$_4$ bilayers. (a) Geometric structures of MnBi$_2$Te$_4$-2L/In$_2$Se$_3$ heterostructures. Left panel shows the top view of the lowest energy configuration. Middle and right panels show the side view of the structures with different polarizations. The thin purple arrows denote spins of the Mn ions. While the thick blue arrows denote polarizations of the FE substrate. (b, c) Band structures for the two states in (a), respectively, i.e., MnBi$_2$Te$_4$-2L(AFM)/In$_2$Se$_3(\uparrow)$ and MnBi$_2$Te$_4$-2L(FM)/In$_2$Se$_3(\downarrow)$. (d) Planar-averaged differential charge density $\Delta \rho(z)$ for the two states shown in (b) and (c). The insets show the density contour at 0.00015 $e$/\AA$^3$. Here, abbreviations (MBT-2L(AFM)/IS($\uparrow$) and MBT-2L(FM)/IS($\downarrow$)) are used by incorporating the IMCs of the MnBi$_2$Te$_4$-2L and the polarization states of In$_2$Se$_3$ for simplicity. }
 \label{fig2}
\end{figure}

We now come to first-principles calculations of the heterostructures of MAT thin films and 2D FE materials. We choose In$_2$Se$_3$ monolayer as the substrate, which has been experimentally proved since its prediction in 2017~\cite{Ding2017,Zhou2017,Cui2018}. We applied the same way for the structural modelings of the heterostructures as Ref.~\onlinecite{Xue2020}, that is, the lattice constants of MAT thin fimls are fixed, whereas that of In$_2$Se$_3$ is adjusted to match those of the overlayers. We have performed careful calculations over a number of stacking configurations (see Fig.~S3 and Table S2~\cite{SM} ). The lowest energy configuration is shown in Fig.~\ref{fig2}a, in which the interfacial Se and Te are in the hollow sites. The stacking order is the same as the one for MnBi$_2$Te$_4$ monolayer on In$_2$Se$_3$~\cite{Xue2020}, which shows up for all MAT bilayers on  In$_2$Se$_3$.
Table~\ref{table1} summarizes the stability of the two magnetic states for MAT bilayers on In$_2$Se$_3$ monolayer in different polarization states.
One can see that for all the MAT bilayers the IMCs remain AFM when the polarization is pointing toward the interface, but become FM as the polarization is reversed. The trend is independent of the vdW functionals/methods (Table S3~\cite{SM}).

\begin{table}[t]
\renewcommand{\arraystretch}{1.25}
\centering
\caption{Energy difference between the interlayer FM and AFM states for freestanding MAT bilayers and their bilayers supported by In$_2$Se$_3$ monolayer in different polarization states (denoted by arrows).
$\Delta E$ = $E_{FM}$ $-$ $E_{AFM}$, $E_{FM}$ ($E_{AFM}$) represents the total energy of the FM (AFM) state. }
\label{table1}
\begin{tabular}{p{3.8cm}<{\centering}p{3.2cm}<{\centering}p{1.2cm}<{\centering}}
\toprule[0.7 pt]
\toprule[0.7 pt]
       Systems & $\Delta E$ [meV/(Mn pair)] & IMCs \\
        \hline
        MnBi$_2$Te$_4$-2L                               &   0.21         &  AFM             \\
        MnBi$_2$Te$_4$-2L/In$_2$Se$_3$($\uparrow$)      &   0.22         &  AFM             \\
        MnBi$_2$Te$_4$-2L/In$_2$Se$_3$($\downarrow$)    &  -0.16         &   FM             \\
        \hline
        MnSb$_2$Te$_4$-2L                               &   0.39         &  AFM               \\
        MnSb$_2$Te$_4$-2L/In$_2$Se$_3$($\uparrow$)      &   0.36         &  AFM             \\
        MnSb$_2$Te$_4$-2L/In$_2$Se$_3$($\downarrow$)    &  -0.40         &   FM             \\
        \hline
        MnBi$_4$Te$_7$-2L                               &   0.03         &  AFM             \\
        MnBi$_4$Te$_7$-2L/In$_2$Se$_3$($\uparrow$)      &   0.03         &  AFM             \\
        MnBi$_4$Te$_7$-2L/In$_2$Se$_3$($\downarrow$)    &  -0.01         &   FM             \\
        \hline
        MnSb$_4$Te$_7$-2L                               &   0.06         &  AFM             \\
        MnSb$_4$Te$_7$-2L/In$_2$Se$_3$($\uparrow$)      &   0.09         &  AFM             \\
        MnSb$_4$Te$_7$-2L/In$_2$Se$_3$($\downarrow$)    &  -0.04         &   FM             \\
\toprule[0.7 pt]
\toprule[0.7 pt]
\end{tabular}
\label{table1}
\end{table}

Below we focus on the electronic structure of MnBi$_2$Te$_4$ bilayer on In$_2$Se$_3$ monolayer, i.e., MnBi$_2$Te$_4$-2L/In$_2$Se$_3$, which are shown in Figs.~\ref{fig2}b-d. Those for all other MAT systems are shown in Figs.~S4$-$S6 since they show pretty much the same trend as MnBi$_2$Te$_4$-2L/In$_2$Se$_3$.
The Fermi level is located in the band gap for the AFM state. Whereas for the FM state, the conduction band of In$_2$Se$_3$ is shifted down into the valence band of MnBi$_2$Te$_4$-2L such that the Fermi level is crossing the valence band of the latter. This feature favors interfacial charge transfer. Fig.~\ref{fig2}d depicts the differential charge density for the two magnetic states, which indicates that there is almost negligible interfacial charge transfer between the MnBi$_2$Te$_4$-2L and In$_2$Se$_3$ for the AFM state. In contrast, the charge transfer is much more significant for the FM state than that for the AFM state. A close inspection finds that the charge density on the Mn atoms in the
interfacial layer becomes positive. This confirms the picture of hole doping over this layer and opens up the $e_g^{\uparrow}-\{p\dots p\}-e_g^{\uparrow}$ hopping channel. Consequently, the FM state becomes energetically favorable for this type of band structure. Thus, the FE In$_2$Se$_3$ monolayer fits the criterion for a substrate that gives switchable band alignments between type-II and type-III with MnBi$_2$Te$_4$-2L. The switchable band alignments induced by a monolayer of FE In$_2$Se$_3$ are robust and unchanged by the use of the HSE06 hybrid functional in the calculations (see Figs.~S7\cite{SM}). Moreover, the trend that the charge transfer mainly happened to the interfacial layer also suggests that the spin-flipping accompanied by the AFM-FM transition takes place to the interfacial MnBi$_2$Te$_4$ layer. For the trilayers and quadlayers, our calculations find the same trend in the spin-flipping as the bilayers (Figs.~S8$-$S9\cite{SM}). We further investigate the effect of substrates on the IMCs in the heterojunctions MnBi$_2$Te$_4$-2L/In$_2$Se$_3$. We chose h-BN as the substrate to construct the sandwich heterojunction of h-BN/MnBi$_2$Te$_4$-2L/In$_2$Se$_3$/h-BN since it has been widely used as a capping layer in devices. Our result show that h-BN has minor effect on the energy difference between the interlayer FM and AFM states (Table S4\cite{SM}), as well as the band alignment between MnBi$_2$Te$_4$-2L and In$_2$Se$_3$ (Fig. S10~\cite{SM}).

\subsection{Magnetic multistates in MAT multilayers}

\begin{figure*}[t]
  \includegraphics[width=.95\linewidth]{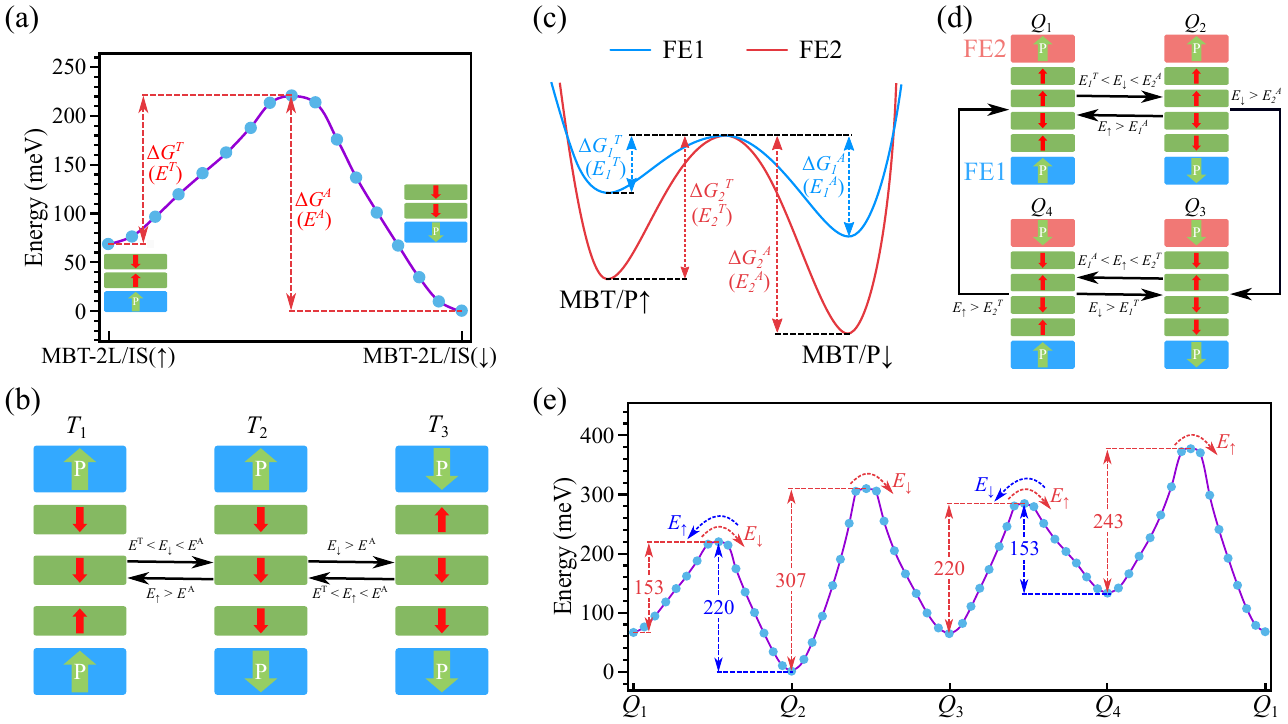}
  \caption{Magnetic multistates in MnBi$_2$Te$_4$ thin films. (a) Kinetic pathway of the FE phase transforming in MnBi$_2$Te$_4$-2L/In$_2$Se$_3$ (abbreviated as MBT-2L/IS). Interface effects lead to asymmetric barrier heights for the two polarization states, which are labelled as $\Delta G^T$ and $\Delta G^A$ as the polarization point toward and away from the interface, respectively. Correspondingly, the critical electric fields are labelled as $E^T$ and $E^A$, respectively. (b) Triple magnetic states in In$_2$Se$_3$/MnBi$_2$Te$_4$-3L/In$_2$Se$_3$ and schematic FE transforming by controlling the external electric field.  $E_{\uparrow}$ ($E_{\downarrow}$) represents the external electric fields along the z (-z) axis. (c) Requirement of energy barriers of the two different FE layers for quadruple magnetic states in sandwich structure FE1/MBT-4L/FE2, $\Delta G_1^T < \Delta G_1^A < \Delta G_2^T < \Delta G_2^A$. $\Delta G_1^T$ and $\Delta G_1^A$ are for one FE layer (FE1), which is colored in blue. Critical electric fields needed to overcome these barriers are denoted as $E_1^T$ and $E_1^A$,respectively. $\Delta G_2^T$ and $\Delta G_2^A$ are for the other layer (FE2) colored in red, for which the critical fields are $E_2^T$ and $E_2^A$, respectively. (d) Schematic illustration of quadruple magnetic states in FE1/MnBi$_2$Te$_4$-4L/FE2 and transforming between the states under electric fields. (e) Kinetic pathways of the quadruple states in In$_2$SSe$_2$/MBT-4L/In$_2$Se$_3$ during FE transforming. The convention of labeling spins of the Mn$^{+2}$ ions and the polarizations of In$_2$Se$_3$ is the same as in Fig.~\ref{fig2}.}
 \label{fig3}
\end{figure*}

The interface has a significant impact on the polarization states of the FE In$_2$Se$_3$ monolayer by introducing a coupling between its polarizations and the local dipoles of MAT. This coupling breaks the symmetry of the two polarization states, that is, it gives rise to asymmetric barrier heights for the two polarization states. Fig.~\ref{fig3}a shows that the state with the polarizations pointing toward the MnBi$_2$Te$_4$ bilayer has a barrier height of about 152 meV ($\Delta G^T$), which is about 68 meV lower than the one with polarizations pointing away from the interface ($\Delta G^A$). The barriers shown in Fig.~\ref{fig3}a larger than that reported by Ref.~\onlinecite{Ding2017} is due to the strain effect (see Fig. S11~\cite{SM}). Nevertheless, our results reveal asymmetry in the energy barriers. Therefore, the critical electric fields needed to flip the polarizations pointing toward the interface ($E^T$) is smaller than that for the reverse process ($E^A$), i.e., $E^T < E^A$. In Fig.~\ref{fig3}a, we assume that the initial AFM ordering in the MnBi$_2$Te$_4$ bilayer is head-to-head. This configuration has almost the same energy as the one with tail-to-tail magnetizations in the case that magnetic anisotropy is not included (The magnetic anisotropy is pretty small, see Ref.~\onlinecite{Xue2020}).

The asymmetric barrier heights along with the unique polarization-dependent IMCs allow designing ferroelectrically switchable magnetic multistates for MAT multilayers. We illustrate the concept in MnBi$_2$Te$_4$ trilayers and quadlayers, i.e., MnBi$_2$Te$_4$-3L and MnBi$_2$Te$_4$-4L. We first sandwich MnBi$_2$Te$_4$-3L in between two In$_2$Se$_3$ layers (Fig.~\ref{fig3}b). Suppose that both the top and bottom In$_2$Se$_3$
layers have up polarizations, which can be achieved by applying external electric fields anyway. According to the polarization dependent IMCs discussed above, spins in the MnBi$_2$Te$_4$ layer next to the top In$_2$Se$_3$ layer will be flipped so that it will be ferromagnetically coupled with the underneath MnBi$_2$Te$_4$ layer. We
label this magnetic state as $T_1$. Then one can apply an electrical field $E_{\downarrow}$ antiparallel to the $z$ axis that is larger than the critical field overcoming $\Delta G^T$ but smaller than the one required to overcome $\Delta G^A$, i.e.,
$E^T < E_{\downarrow} < E^A$. As a result, the polarizations in the bottom layer will be reversed while those in the top layer will remain unchanged. Then, the magnetization of the bottom MnBi$_2$Te$_4$ layer will be flipped to be ferromagnetically coupled with the adjacent MnBi$_2$Te$_4$ layer, i.e., $T_2$ in Fig.~\ref{fig3}b. Further increasing the electric field such that
$E^{\downarrow} > E^A$ will also drive the polarizations of the top In$_2$Se$_3$ layer to be flipped. Correspondingly, the magnetizations of the top MnBi$_2$Te$_4$ layer will be flipped,
for which the magnetic state is labelled as $T_3$. Now, an electric field along the $z$ axis, i.e., $E^{\uparrow}$, will first force the polarization of the bottom In$_2$Se$_3$ to be reversed when $E^T < E^{\uparrow} < E^A$. As a result, the system will flow into $T_2$. Further enhancing $E^{\uparrow}$ to the level that $E^{\uparrow} > E^A$ will drive the system back into $T_1$.
So the whole system have triple magnetic states, which can be ferroelectrically controlled. Likewise, sandwiching thicker films than triple layers by the same FE layers also gives rise to triple magnetic states.

More magnetic states can be obtained by sandwiching the MAT thin films in between two different FE layers with a special combination of the barrier heights. Such a combination requires that the highest barrier for one FE monolayer should be lower than the lowest barrier for the other FE layer. We depict the barrier heights for the two different FE layers in  Fig\ref{fig3}c, $\Delta G_1^T$ and $\Delta G_1^A$ are for one FE layer (FE1), to which the corresponding critical electric fields are $E_1^T$ and $E_1^A$, respectively. Whereas $\Delta G_2^T$ and $\Delta G_2^A$ are for the other layer colored in red (FE2), for which the critical fields are $E_2^T$ and $E_2^A$, respectively. In the case that $\Delta G_1^T < \Delta G_1^A < \Delta G_2^T < \Delta G_2^A$, i.e., $E_1^T < E_1^A < E_2^T < E_2^A$, a layer-by-layer flipping mechanism for the FE contacts can be achieved by properly controlling the electric field. As a result, one can have quadruple magnetic states based on the polarization-dependent IMCs in MAT heterostructures (Fig.~\ref{fig3}d). Our calculations find that the barrier heights of In$_2$Se$_3$ and In$_2$SSe$_2$ monolayers fit the above requirement for the quadruple magnetic states. Specifically, we obtain 245 meV ($\Delta G_2^T$) and 308 meV ($\Delta G_2^A$) for In$_2$SSe$_2$ with polarizations pointing toward and away from the MnBi$_2$Te$_4$ layer (see Figs. S11 and S12~\cite{SM}), respectively, which are larger than those of In$_2$Se$_3$ (see Fig.~\ref{fig3}a, 152 meV for $\Delta G_1^T$ and 220 meV for $\Delta G_1^A$). In Fig.~\ref{fig3}e, we show the kinetic pathway of transforming the polarization states, which suggests that the quadruple states are ferroelectrically switchable.

\subsection{Topological properties of MAT thin films}
The ferroelectrically tunable magnetic multistates give rise to a variety of distinct topological properties for the MAT thin films. In MAT systems, each helical surface state contribute half-quantized Hall conductance. So, the Chern number ($C$) will be 0 when magnetizations of the two surfaces of the thin film are antiparallel, and will be 1/-1 when the surface magnetizations are parallel with up/down spin-polarizations.  Therefore, according to our results shown in  Fig.~\ref{fig3}, there are two different Chern numbers for In$_2$Se$_3$/MnBi$_2$Te$_4$-3L/In$_2$Se$_3$ and In$_2$Se$_3$/MnBi$_2$Te$_4$-4L/In$_2$Se$_3$, and three Chern numbers for In$_2$SSe$_2$/MnBi$_2$Te$_4$-4L/In$_2$Se$_3$, respectively. We perform calculations of the Chern number  for the
MnBi$_2$Te$_4$ multilayers with the magnetic states shown in Figs.~\ref{fig2} and ~\ref{fig3}. For the bilayer systems, the topological properties of MnBi$_2$Te$_4$ remain unchanged upon interfacing,
i.e., $C = 1$ for the FM state and $C = 0$ for the AFM state, which is also supported by the results of edge states (Figs.~S13 and S14). For MnBi$_2$Te$_4$-3L, we have $C$ = 0, -1, and 0 for $T_1$,
$T_2$, and $T_3$, respectively. Whereas for MnBi$_2$Te$_4$-4L, there are three Chern numbers for the quadruple states, i.e., 1, 0, and -1. Table~\ref{table2} summarizes the Chern numbers for the studied MnBi$_2$Te$_4$ thin films. We expect that such a ferroelectrically tunable multiplet for the Chern number may be seen in other MAT multilayers.

\begin{table}[t]
\renewcommand{\arraystretch}{1.25}
\centering
\caption{Chern number ($C$) for MnBi$_2$Te$_4$ multilayers with different magnetic states. Arrows denote the magnetizations on Mn ions. }
\label{table2}
\begin{tabular}{p{2.8cm}<{\centering}p{2.2cm}<{\centering}p{1.2cm}<{\centering}}
\toprule[0.7 pt]
\toprule[0.7 pt]
       Systems & IMCs & $C$ \\
        \hline
        \multirow {2}*{MnBi$_2$Te$_4$-2L}       &    M$_1$ ($\uparrow\downarrow$)                        &    0            \\
              ~                                 &    M$_2$ ($\downarrow\downarrow$)                      &   -1            \\
        \hline
        \hline
        \multirow {3}*{MnBi$_2$Te$_4$-3L}       &    T$_1$ ($\downarrow\downarrow\uparrow$)              &    0            \\
              ~                                 &    T$_2$ ($\downarrow\downarrow\downarrow$)            &   -1            \\
              ~                                 &    T$_3$ ($\uparrow\downarrow\downarrow$)              &    0            \\
        \hline
        \hline
        \multirow {4}*{MnBi$_2$Te$_4$-4L}       &    Q$_1$ ($\uparrow\uparrow\downarrow\uparrow$)        &    1            \\
              ~                                 &    Q$_2$ ($\uparrow\uparrow\downarrow\downarrow$)      &    0            \\
              ~                                 &    Q$_3$ ($\downarrow\uparrow\downarrow\downarrow$)    &   -1            \\
              ~                                 &    Q$_4$ ($\downarrow\uparrow\downarrow\uparrow$)      &    0            \\
        \hline
\toprule[0.7 pt]
\toprule[0.7 pt]
\end{tabular}
\label{table2}
\end{table}

In conclusion, we have proposed to ferroelectrically tune the magnetism of MAT thin films using model and first-principles calculations. The scheme is based on the fact
that the IMCs are strongly dependent on the occupation of $d$-orbitals of the Mn$^{2+}$ ions. The
variation in the occupation can be controlled by interfacing the films with a FE layer with appropriate band alignments.  We have demonstrated the concept in MAT/In$_2$Se$_3$
heterostructures by performing first-principles calculations. We find that the interfacing effect mainly has an impact on the interfacial MAT layer.
Specifically, there is spin-flipping in the interfacial layer when polarizations of the In$_2$Se$_3$ are reversed, which results in ferroelectrically switchable IMCs and an AFM-to-FM transition. On the other hand, the
interfacing effect leads to asymmetric energy barrier heights, which means that different electric fields are needed to switch the polarizations for the two states. We
further show that this physics can be used to build magnetic multistates in their sandwich structures. Our calculations suggest that triple and quadruple magnetic states with tunable Chern number can be obtained for MnBi$_2$Te$_4$ thin films by sandwiching them in between appropriate FE layers.  Our results will not only attract experimental interest in FE control of the magnetism and topological properties of MAT thin films, but also inspire designing novel magnetism in other 2D materials.

\begin{acknowledgments}
We thank Haijun Zhang and Zhixin Guo for useful discussions. This work was supported by the National Natural Science Foundation of China (Grants No. 12174098, No. 11774084, No. U19A2090 and No. 91833302) and project supported by State Key Laboratory of Powder Metallurgy, Central South University, Changsha, China.
\end {acknowledgments}

\bibliography{references}
\bibliographystyle{apsrev4-2}

\widetext
\clearpage
\begin{center}
\textbf{\LARGE Supplementary Materials for “Ferroelectrically switchable magnetic multistates in MnBi$_2$Te$_4$(Bi$_2$Te$_3$)$_n$ and MnSb$_2$Te$_4$(Sb$_2$Te$_3$)$_n$ ($n$ = 0 and 1) thin films”}
\end{center}

\setcounter{equation}{0}
\setcounter{figure}{0}
\setcounter{table}{0}
\setcounter{page}{1}
\makeatletter
\renewcommand{\theequation}{S\arabic{equation}}
\renewcommand{\thefigure}{S\arabic{figure}}
\renewcommand{\bibnumfmt}[1]{[S#1]}
\renewcommand{\citenumfont}[1]{S#1}
~\\
~\\

\setcounter{section}{0}

\section*{\Large 1. Doping induced AFM-to-FM IMC transition in MnBi$_4$Te$_7$ and MnSb$_4$Te$_7$ bilayers}

\begin{figure}[htbp]
\centering
\includegraphics[width=0.55\textwidth]{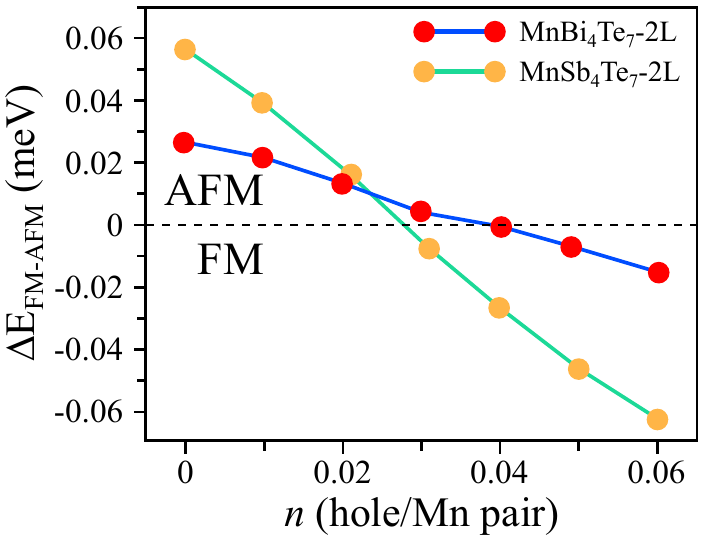}
  \caption{Energy difference between the FM and AFM states as a function of hole doping, i.e., $\Delta E = E_{FM} - E_{AFM}$.}
\label{S1}
\end{figure}

\newpage
\section*{\Large 2. Magnetic exchange constants}

\begin{figure}[htbp]
\centering
\includegraphics[width=0.55\textwidth]{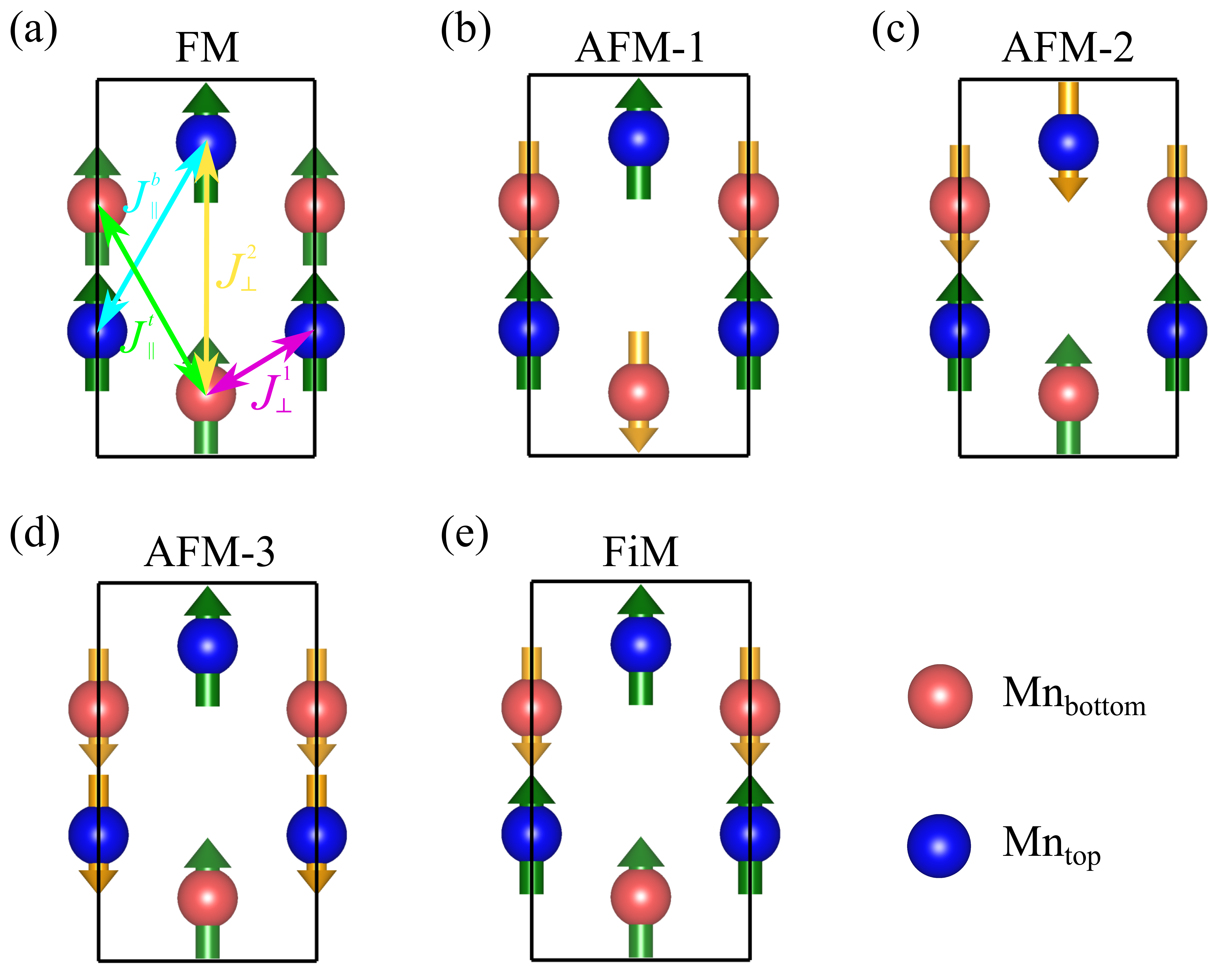}
  \caption{Five spin configurations for calculating the exchange parameters of MnA$_2$Te$_4$(A$_2$Te$_3$)$_n$ bilayer (MAT-2L). Here, only the Mn atoms are shown.  The Mn ions in different MAT layers are in different colors. (a) Ferromagnetic configuration (FM). $J_{\parallel}^{t}$ (green arrow) and $J_{\parallel}^{b}$ (cyan arrow) represent the first nearest-neighbor (1st) intralayer exchange interactions between the Mn ions in the top and bottom layers, respectively.  $J_{\parallel}^{t} = J_{\parallel}^{b}$ for freestanding bilayers. The interlayer exchange interaction for the first and second nearest-neighbors ($J_{\perp}^{1st}$ and $J_{\perp}^{2nd}$) are indicated by the magenta and yellow arrows, respectively. (b-e) Antiferromagnetic-1, 2, and 3 configurations, which are denoted as AFM-1,  AFM-2, and AFM-3, respectively. (e) Ferrimagnetic configuration (FiM). The arrows on the Mn ions denote the spins.}
\label{S2}
\end{figure}

To determine the intra- and interlayer exchange couplings between the Mn ions, we constructed a $1 \times \sqrt{3}$ supercell and mapped the DFT total energies to the Hamiltonian of five spin configurations (see Fig.~\ref{S2}). The total energies of the considered spin configurations can be written as:

\begin{equation}
\begin{split}
\begin{aligned}
&&E_{FM}&=E_0 + (12J_{\parallel}^{t} + 12J_{\parallel}^{b} + 12J_{\perp}^{1st} + 12J_{\perp}^{2nd})S^2 &\\
&&E_{AFM-1}&=E_0 + (12J_{\parallel}^{t} + 12J_{\parallel}^{b} - 12J_{\perp}^{1st} - 12J_{\perp}^{2nd})S^2 &\\
&&E_{AFM-2}&=E_0 + (-4J_{\parallel}^{t} - 4J_{\parallel}^{b} + 4J_{\perp}^{1st} - 12J_{\perp}^{2nd})S^2 &\\
&&E_{AFM-3}&=E_0 + (-4J_{\parallel}^{t} - 4J_{\parallel}^{b} - 4J_{\perp}^{1st} + 12J_{\perp}^{2nd})S^2 &\\
&&E_{iFM}&=E_0 + (-4J_{\parallel}^{t} + 12J_{\parallel}^{b})S^2 &
\end{aligned}
\end{split}
\end{equation}
The derived magnetic exchange constants are summarized in Table~\ref{table1}.

\begin{table}[h]
\renewcommand{\arraystretch}{1.25}
\centering
\caption{Doping dependence of $J_{\perp}^{1st}$, $J_{\perp}^{2nd}$, and $\bar{J_{\perp}}$ (in meV).}
\label{table1}
\begin{tabular}{p{1.8cm}<{\centering}p{1.2cm}<{\centering}p{1.2cm}<{\centering}p{1.2cm}<{\centering}p{1.2cm}<{\centering}p{1.2cm}<{\centering}p{1.2cm}<{\centering}p{1.2cm}<{\centering}p{1.2cm}<{\centering}}
\toprule[0.7 pt]
\toprule[0.7 pt]
           Systems       & $J$     &    0.00    &   0.01   &   0.02    &   0.03    &   0.04    &   0.05    &   0.06  \\
        \hline
        \multirow{3}*{MnBi$_2$Te$_4$-2L} &   $J_{\perp}^{1st}$ &    ~0.045    &   ~0.037    &    ~0.027     &    ~0.014     &    -0.004    &    -0.023     &    -0.041     \\
        ~ &  $J_{\perp}^{2nd}$ &    -0.010     &    -0.011    &    -0.013     &    -0.017     &    -0.021    &    -0.029    &    -0.040     \\
        ~ &  $\bar{J_{\perp}}$ &    ~0.035    &    ~0.026   &    ~0.014     &    -0.003     &    -0.025   &    -0.052    &    -0.081     \\
        \hline
        \multirow{3}*{MnSb$_2$Te$_4$-2L} &   $J_{\perp}^{1st}$ &    ~0.075     &    ~0.053    &    ~0.031     &    ~0.008    &    -0.014   &    -0.034     &    -0.051     \\
        ~ &  $J_{\perp}^{2nd}$ &    -0.010     &    -0.016    &    -0.021     &    -0.027     &    -0.032    &    -0.039     &    -0.047     \\
        ~ &  $\bar{J_{\perp}}$ &    ~0.065    &    ~0.037   &    ~0.010    &    -0.019     &   -0.046    &    -0.073     &    -0.098     \\
        \hline
        \hline
\toprule[0.7 pt]
\toprule[0.7 pt]
\end{tabular}
\end{table}

\newpage
\section*{\Large 3. Geometric structures and energies of MAT-2L/In$_2$Se$_3$}

\begin{figure}[htbp]
\centering
\includegraphics[width=0.8\textwidth]{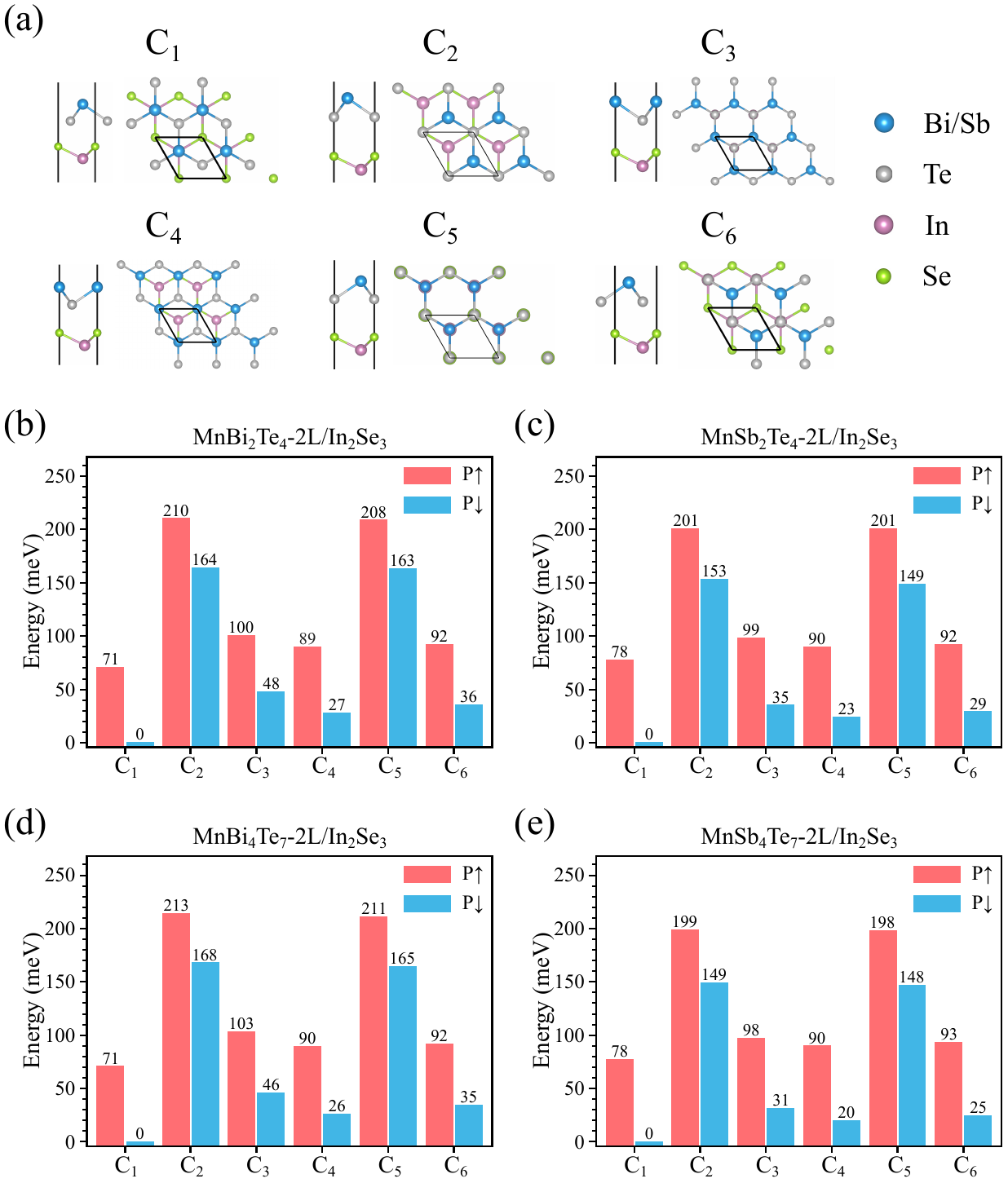}
  \caption{Geometric structures and energetics of MAT-2L/In$_2$Se$_3$ with different polarization states and stackings.
  (a) Geometry of MAT-2L/In$_2$Se$_3$ heterostructures. Here, only the four atomic layers (Bi-Te/Se-In) at the interface
  are shown.  (b - e) Energie of MnBi$_2$Te$_4$-2L/In$_2$Se$_3$,  MnSb$_2$Te$_4$-2L/In$_2$Se$_3$,
  MnBi$_4$Te$_7$-2L/In$_2$Se$_3$, and  MnSb$_4$Te$_7$-2L/In$_2$Se$_3$ with different polarization states. P($\uparrow$) and P($\downarrow$)
  represents the polarizations of In$_2$Se$_3$ pointing toward and away from the interface, respectively.}
\label{S3}
\end{figure}

We considered six stacking configurations for MAT-2L/In$_2$Se$_3$ heterostructures.
Fig.~\ref{S3}(a) shows the side and top views of the heterostructures, for which only the four atomic layers at the interface (Bi-Te/Se-In)
are displayed. The energies are shown in Figs.~\ref{S3}(b)-(e), for which that of configuration C$_1$ with downward polarization (pointing away from the interface) is taken as the reference.
For all the configurations, the state with the polarization pointing toward MAT has a higher energy than the one with the
polarization pointing away from the interface. This trend can be understood with the interlayer distance differences between them (see Table.~\ref{table2}).

\newpage
\section*{\Large 4. Interlayer distance between MAT-2L and In$_2$Se$_3$ in different stacking configurations}

The interlayer distance ($d$ as denoted in Fig. 2 and Figs.~\ref{S4}-\ref{S6}) between MAT-2L and In$_2$Se$_3$ in their
heterostructures are summarized in Table.~\ref{table2}.  One
can see that for each configuration the state with downward polarization (P$\downarrow$) has a smaller distance than that with upward polarization (P$\uparrow$).
This result indicates that the former has stronger interlayer coupling between MAT-2L and In$_2$Se$_3$.
Therefore, the state with the polarizations pointing away MAT-2L has lower energy than polarizations
pointing toward the interface (see Fig.~\ref{S3}). Another trend in the interlayer distance is that the C$_1$ configuration has the smallest
distance compared to other stacking configurations.


\begin{table}[h]
\renewcommand{\arraystretch}{1.16}
\centering
\caption{Interlayer distances between MAT-2L and In$_2$Se$_3$ in the their heterostructures.
C$_i$ represent the stacking configurations. Interlayer distances are shown in \AA.}
\label{table2}
\begin{tabular}{p{4cm}<{\centering}p{1.42cm}<{\centering}p{1.42cm}<{\centering}p{1.42cm}<{\centering}p{1.42cm}<{\centering}p{1.42cm}<{\centering}p{1.42cm}<{\centering}}
\toprule[0.7 pt]
\toprule[0.7 pt]
           Systems       & C$_1$    & C$_2$  & C$_3$  & C$_4$  & C$_5$  & C$_6$  \\
        \hline
        MnBi$_2$Te$_4$-2L/In$_2$Se$_3$($\uparrow$)    &    2.82     &    3.79    &    3.13     &    2.93     &    3.87    &    3.01     \\
        MnBi$_2$Te$_4$-2L/In$_2$Se$_3$($\downarrow$)  &    2.60     &    3.64    &    2.96     &    2.78     &    3.64    &    2.89     \\
        \hline
        MnSb$_2$Te$_4$-2L/In$_2$Se$_3$($\uparrow$)    &    3.00     &    3.79    &    3.20     &    3.01     &    3.79    &    3.13     \\
        MnSb$_2$Te$_4$-2L/In$_2$Se$_3$($\downarrow$)  &    2.73     &    3.63    &    3.05     &    2.91     &    3.64    &    3.04     \\
        \hline
        MnBi$_4$Te$_7$-2L/In$_2$Se$_3$($\uparrow$)    &    2.83     &    3.85    &    3.09     &    2.93     &    3.82    &    3.01     \\
        MnBi$_4$Te$_7$-2L/In$_2$Se$_3$($\downarrow$)  &    2.62     &    3.67    &    2.99     &    2.80     &    3.65    &    2.90     \\
        \hline
        MnSb$_4$Te$_7$-2L/In$_2$Se$_3$($\uparrow$)    &    2.93     &    3.80    &    3.17     &    3.03     &    3.80    &    3.05     \\
        MnSb$_4$Te$_7$-2L/In$_2$Se$_3$($\downarrow$)  &    2.66     &    3.64    &    3.00     &    2.85     &    3.63    &    2.98     \\
        \hline
\toprule[0.7 pt]
\toprule[0.7 pt]
\end{tabular}
\end{table}

\section*{\Large 5. Stability of different interlayer magnetic states for freestanding and supported MAT-2L}

The energies of the interlayer FM and AFM orderings in the main text were calculated using DFT-D3 method of Grimme with zero-damping
function. In addition, we have performed calculations using different vdWs functionals/methods for which the results are summarized in Table.~\ref{table3}.
One can see that all the methods/functionals produce the same trend in the stability.

\begin{table}[h]
\renewcommand{\arraystretch}{1.16}
\centering
\caption{Energy differences ($\Delta E$) between the interlayer FM and AFM couplings for freestanding MAT-2L and MAT-2L/In$_2$Se$_3$. $\Delta E$ is defined as $\Delta E = E_{FM} - E_{AFM}$, where $E_{FM}$ ($E_{AFM}$) represent the total energy of the FM (AFM) state. Different vdWs functionals/methods were used in the total energy calculations for comparison. Here, DFT-D3$^1$ and DFT-D3$^2$ represents the DFT-D3 method of Grimme with zero-damping function and with Becke-Johnson damping function, respectively.}
\label{table3}
\begin{tabular}{p{4cm}<{\centering}p{1.7cm}<{\centering}p{1.7cm}<{\centering}p{1.7cm}<{\centering}p{1.7cm}<{\centering}p{1.7cm}<{\centering}}
\toprule[0.7 pt]
\toprule[0.7 pt]
                    Systems   &   \multicolumn{4}{c}{$\Delta E$ [meV/(Mn pair)]}   &  \multirow{2}*{IMC}                               \\
                    vdW methods        &      DFT-D2     &     DFT-D3$^1$     &      DFT-D3$^2$       &      optPBE      &                 \\
        \hline
        MnBi$_2$Te$_4$-2L                              &     0.23     &     0.21     &     0.27     &     0.29         &  AFM            \\
        MnBi$_2$Te$_4$-2L/In$_2$Se$_3$($\uparrow$)     &     0.23     &     0.22     &     0.22     &     0.28         &  AFM            \\
        MnBi$_2$Te$_4$-2L/In$_2$Se$_3$($\downarrow$)   &    -0.14     &    -0.16     &    -0.07     &    -0.11         &   FM            \\
        \hline
        MnSb$_2$Te$_4$-2L                              &     0.40     &     0.39     &     0.57     &     0.50         &  AFM            \\
        MnSb$_2$Te$_4$-2L/In$_2$Se$_3$($\uparrow$)     &     0.32     &     0.36     &     0.51     &     0.48         &  AFM            \\
        MnSb$_2$Te$_4$-2L/In$_2$Se$_3$($\downarrow$)   &     0.38     &    -0.40     &    -0.55     &    -0.22         &   FM            \\
        \hline
        MnBi$_4$Te$_7$-2L                              &     0.03     &     0.03     &     0.04     &     0.02         &  AFM            \\
        MnBi$_4$Te$_7$-2L/In$_2$Se$_3$($\uparrow$)     &     0.02     &     0.03     &     0.03     &     0.01         &  AFM            \\
        MnBi$_4$Te$_7$-2L/In$_2$Se$_3$($\downarrow$)   &    -0.01     &    -0.01     &    -0.01     &    -0.01         &   FM            \\
        \hline
        MnSb$_4$Te$_7$-2L                              &     0.07     &     0.06     &     0.12     &     0.05         &  AFM            \\
        MnSb$_4$Te$_7$-2L/In$_2$Se$_3$($\uparrow$)     &     0.07     &     0.09     &     0.11     &     0.05         &  AFM            \\
        MnSb$_4$Te$_7$-2L/In$_2$Se$_3$($\downarrow$)   &    -0.06     &    -0.04     &    -0.06     &    -0.03         &   FM            \\

\toprule[0.7 pt]
\toprule[0.7 pt]
\end{tabular}
\end{table}

\newpage
\section*{\Large 6. Electronic structure of MAT-2L/In$_2$Se$_3$}

\begin{figure}[htbp]
\centering
\includegraphics[width=0.73\textwidth]{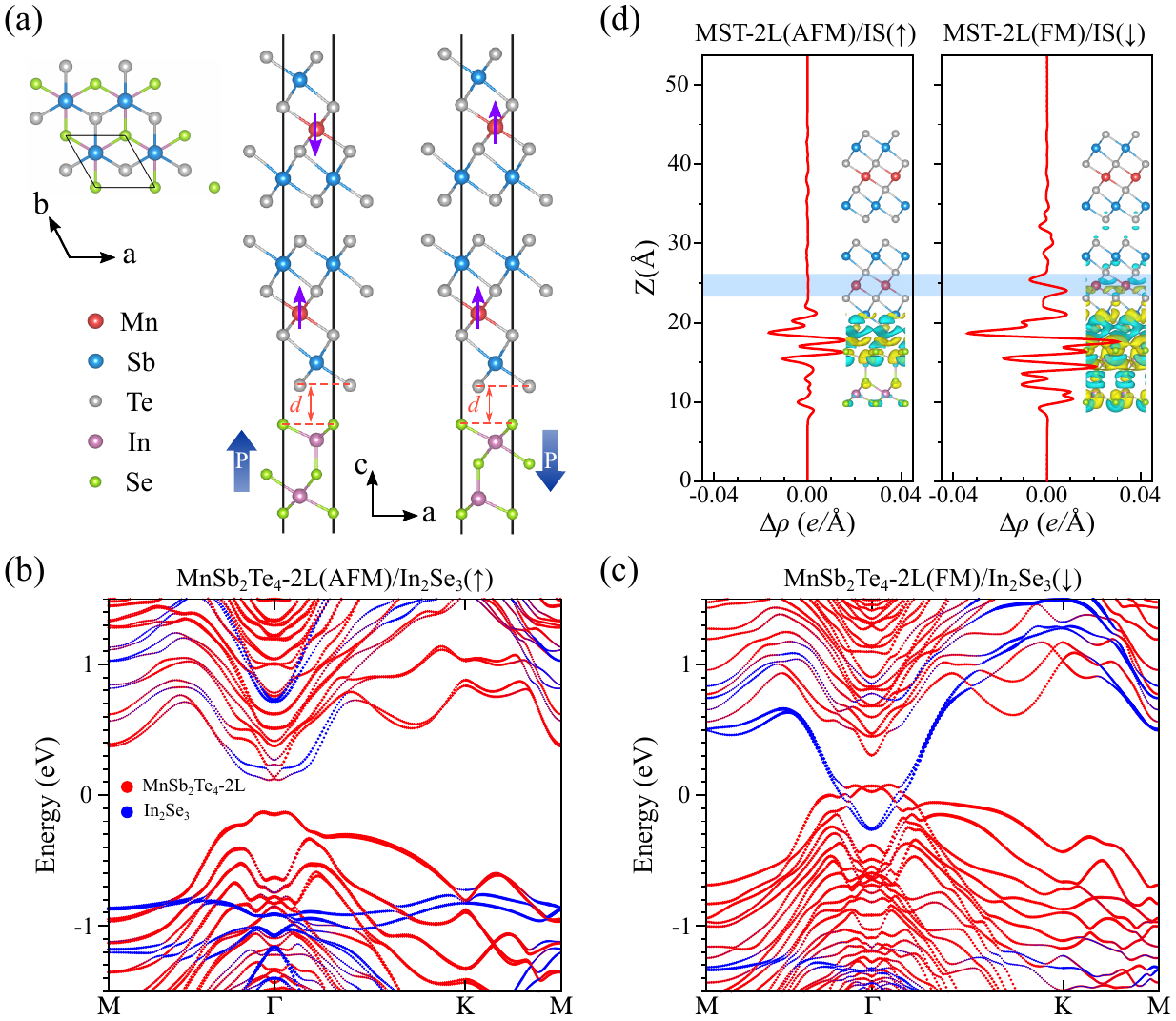}
\caption{Interface effect on the electronic structure of MnSb$_2$Te$_4$ bilayers. (a) Geometric structures of MnSb$_2$Te$_4$-2L/In$_2$Se$_3$.
The purple and blue arrows indicates the magnetic moment orientation of the Mn atoms
and the dipole moment directed of the In$_2$Se$_3$. $d$ represents the interlayer distance between MnSb$_2$Te$_4$-2L and In$_2$Se$_3$.  Layer-projected band structures of (c) MnSb$_2$Te$_4$-2L/In$_2$Se$_3$($\uparrow$) and (d)
MnSb$_2$Te$_4$-2L/In$_2$Se$_3$($\downarrow$) with SOC, respectively. The blue and red lines represents the band structure of MnSb$_2$Te$_4$-2L and In$_2$Se$_3$, respectively. (d) Planar averaged
charge density difference of MnSb$_2$Te$_4$-2L/In$_2$Se$_3$. The insets shows the differential charge density of MnSb$_2$Te$_4$-2L/In$_2$Se$_3$ at the isosurface 0.00015 $e$/Bohr$^3$. The yellow and blue isosurface represents the electrons accumulation and reduction. The charge transfer of
Mn atoms at the interface is marked with shading. }
\label{S4}
\end{figure}

\begin{figure}[htbp]
\centering
\includegraphics[width=0.67\textwidth]{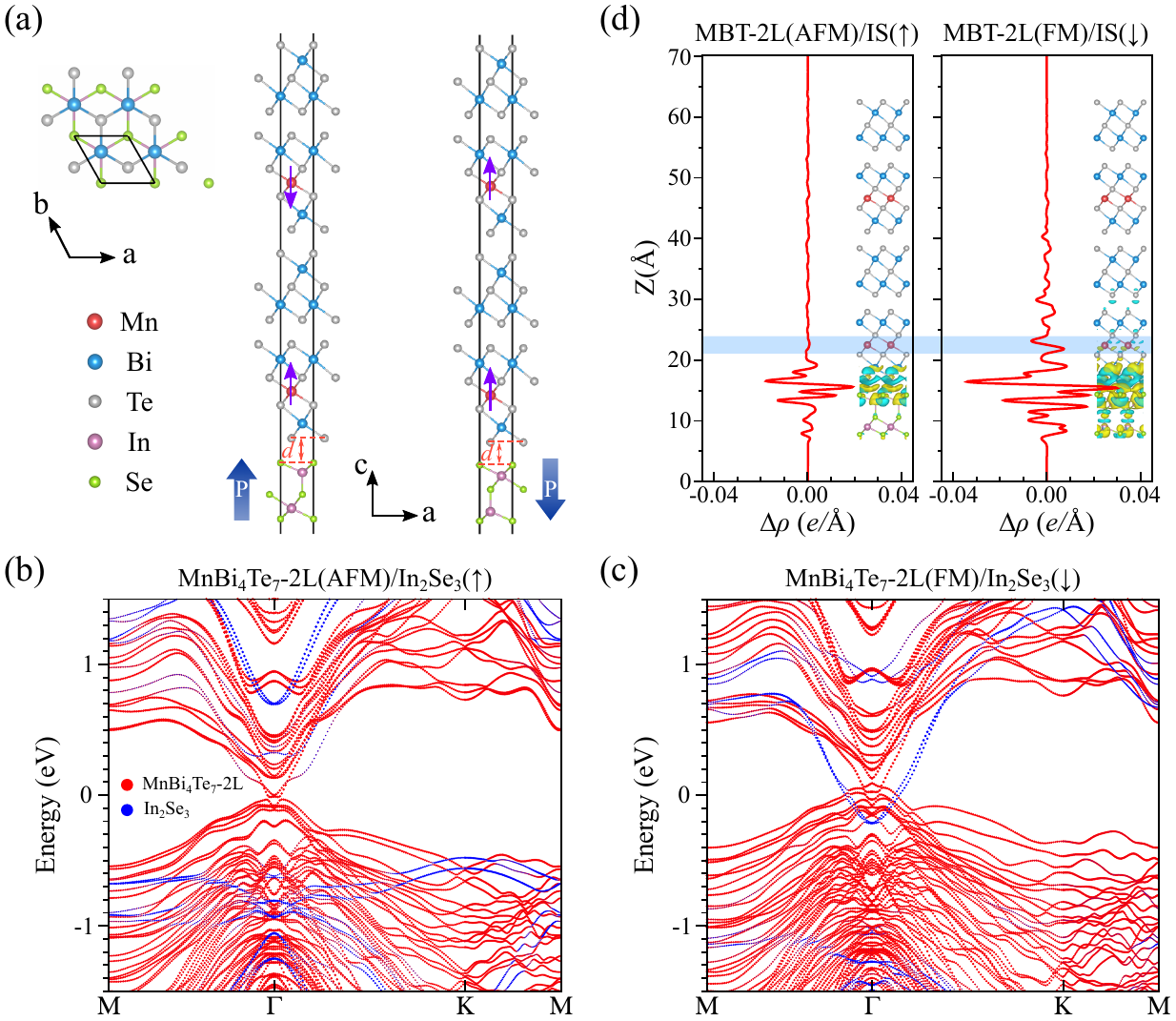}
\caption{Interface effect on the electronic structure of MnBi$_4$Te$_7$ bilayers. The figures are similar as Fig.~\ref{S2}.}
\label{S5}
\end{figure}

\begin{figure}[htbp]
\centering
\includegraphics[width=0.67\textwidth]{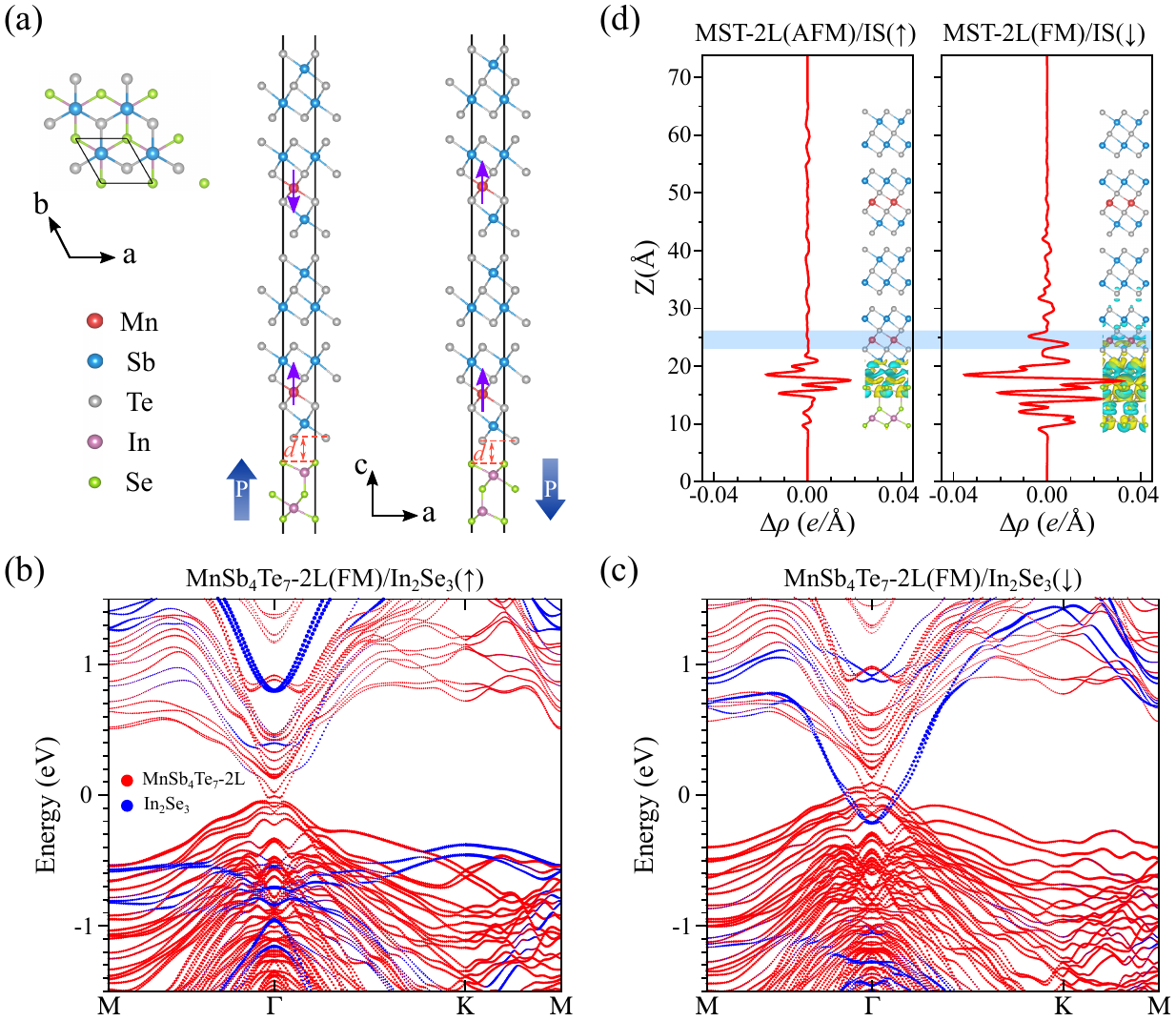}
\caption{Interface effect on the electronic structure of MnSb$_4$Te$_7$ bilayers. The figures are similar as Fig.~\ref{S2}.}
\label{S6}
\end{figure}

Figures~\ref{S4}-\ref{S6} show the geometric structures, layer-projected band structures, and differential charge density of MnSb$_2$Te$_4$-2L/In$_2$Se$_3$,
MnBi$_4$Te$_7$-2L/In$_2$Se$_3$, and MnSb$_4$Te$_7$-2L/In$_2$Se$_3$, respectively.
Similar to MnBi$_2$Te$_4$-2L/In$_2$Se$_3$ systems, the MAT-2L and FE In$_2$Se$_3$ monolayer with different polarization have a type-I and type-III band alignments.
For MAT-2L/In$_2$Se$_3$($\uparrow$), one
can see that charge transfer at the interface is negligible. Therefore, the IMCs remains to be AFM. In contrast, when the polarization flipping
to away from the interface, interlayer charge transfer becomes significant, which gives rise to hole doping to the Mn ions in the interfacial layer.
As a result, the FM state becomes energetically favorable.
~\\
~\\

\newpage

\section*{\Large 7. Band structure of freestanding MnBi$_2$Te$_4$-2L and MnBi$_2$Te$_4$-2L/In$_2$Se$_3$ with HSE06 method}
\begin{figure}[htbp]
\centering
\includegraphics[width=0.8\textwidth]{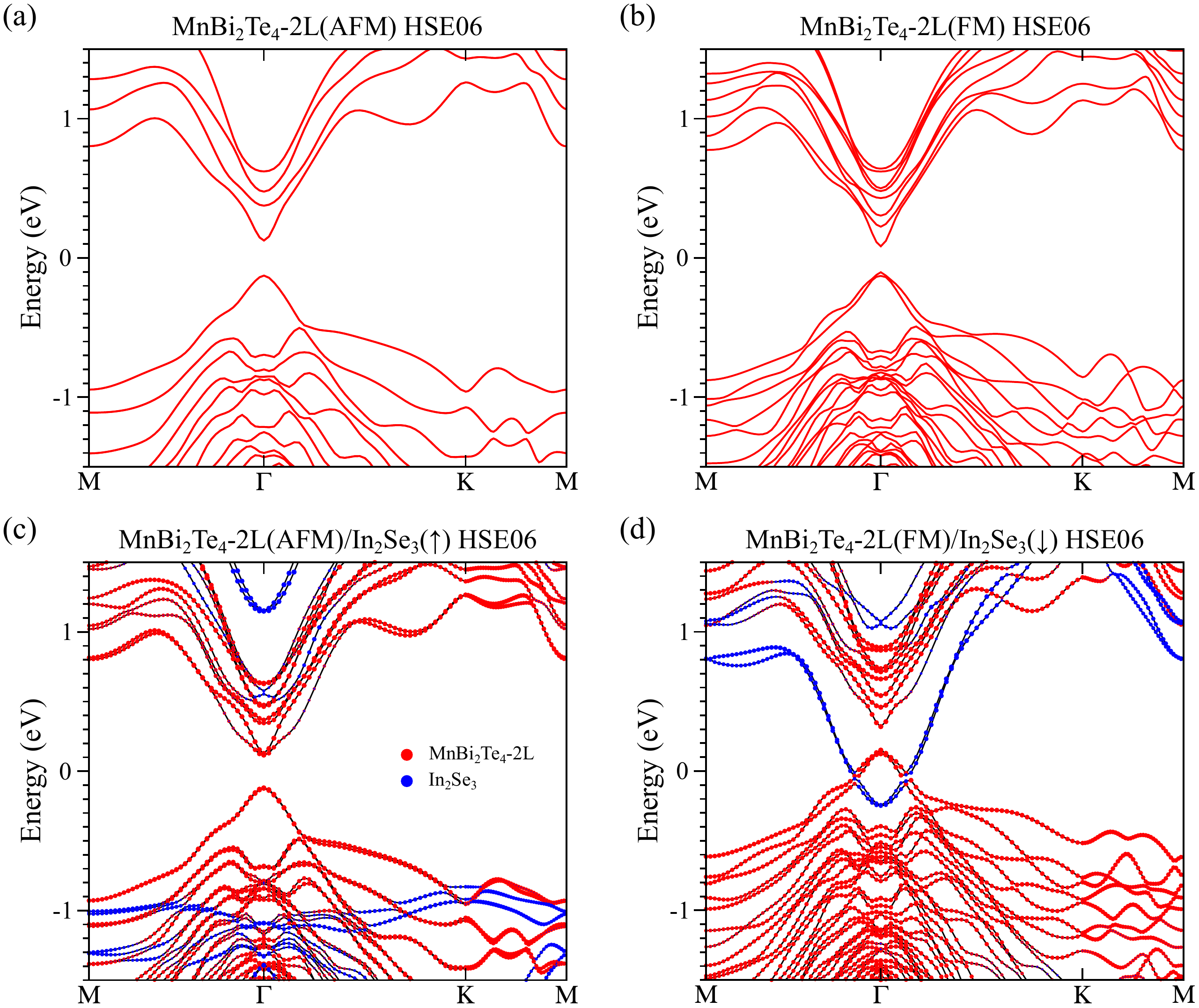}
\caption{Band structure of freestanding MnBi$_2$Te$_4$-2L and MnBi$_2$Te$_4$-2L/In$_2$Se$_3$ with HSE06 method. Band structure of MnBi$_2$Te$_4$-2L with FM and AFM IMC calculated by HSE06 including SOC. Layer-projected band structures of (c) MnBi$_2$Te$_4$-2L/In$_2$Se$_3$($\uparrow$) and (d) MnBi$_2$Te$_4$-2L/In$_2$Se$_3$($\downarrow$) calculated by HSE06 including SOC, respectively. The blue and red lines represents the band structure of MnBi$_2$Te$_4$-2L and In$_2$Se$_3$, respectively.}
\label{S7}
\end{figure}

Figure~\ref{S7} displays the band structures of both freestanding MnBi$_2$Te$_4$-2L and MnBi$_2$Te$_4$-2L/In$_2$Se$_3$, which were computed using the HSE06 method with SOC. Fig~\ref{S7}(a) and (b) illustrate that the FM and AFM states of freestanding MnBi$_2$Te$_4$-2L have direct band gaps of around 186 and 252 meV, respectively. In contrast, the bandgap of 2L-MBT for FM and AFM IMCs was reduced to 170 and 236 meV when forming a heterojunction with In$_2$Se$_3$.  Moreoever, Fig~\ref{S7} further reveals that the HSE06 and PEB methods produced the same trend in terms of the switchable band alignment induced by monolayer FE In$_2$Se$_3$.

\newpage
\section*{\Large 8. Interface effects on the electronic structure of MnBi$_2$Te$_4$ trilayers and quad-layers}
\begin{figure}[htbp]
\centering
\includegraphics[width=0.8\textwidth]{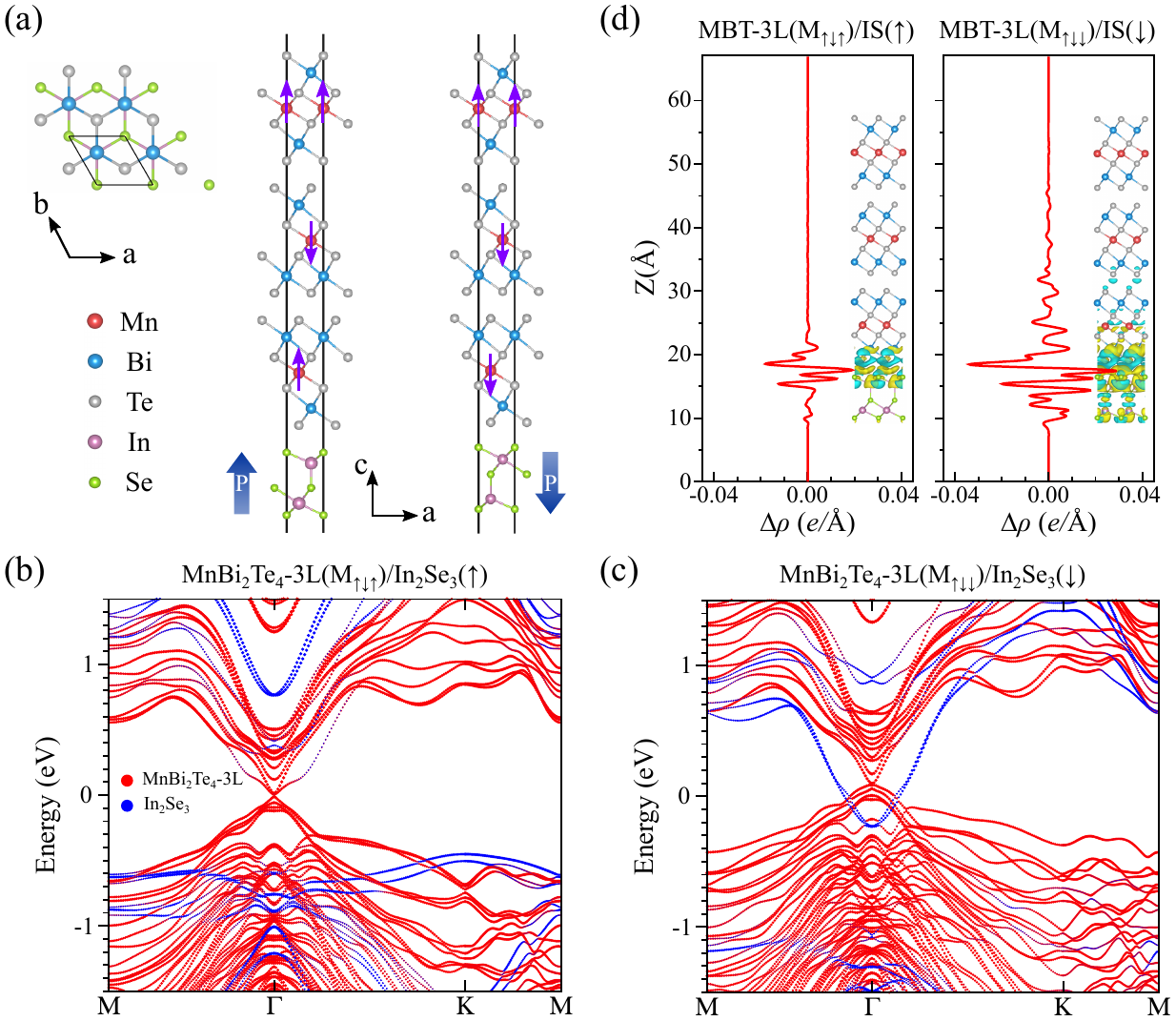}
\caption{Interface effeces on the electronic structure of MnBi$_2$Te$_4$ trilayers. The figures are similar as Fig.~\ref{S2}.}
\label{S8}
\end{figure}

Figures~\ref{S8} and \ref{S9} show the geometric structures, layer-projected band structures, and differential charge density of MnBi$_2$Te$_4$-3L/In$_2$Se$_3$ and MnBi$_2$Te$_4$-4L/In$_2$Se$_3$, respectively.
For the MnBi$_2$Te$_4$ trilayers and quad-layers systems, our calculations find that spin flipping is only taken place to
the interfacial MnBi$_2$Te$_4$ layer. The results are similar to MnBi$_2$Te$_4$-2L/In$_2$Se$_3$.

~\\
~\\

\newpage
\begin{figure}[htbp]
\centering
\includegraphics[width=0.8\textwidth]{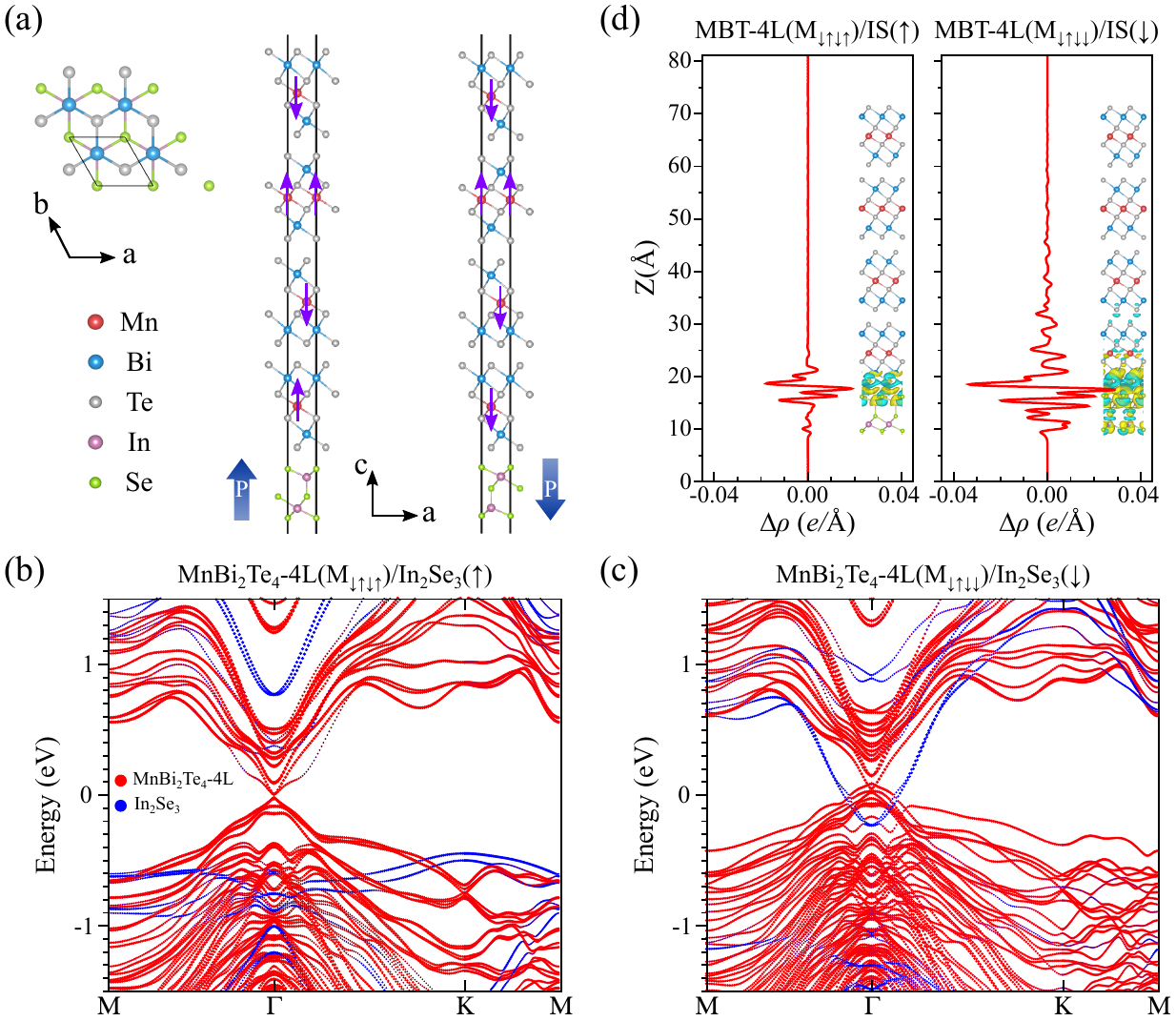}
\caption{Interface effeces on the electronic structure of MnBi$_2$Te$_4$ quad-layers. The figures are similar as Fig.~\ref{S2}.}
\label{S9}
\end{figure}

\newpage

\section*{\Large 9. Effects of substrates on the electronic structure and IMC of MnBi$_2$Te$_4$-2L/In$_2$Se$_3$}

We considered nine different stacking configurations for h-BN/MnBi$_2$Te$_4$-2L/In$_2$Se$_3$/h-BN heterostructures. Fig.~\ref{S9}(a) shows the top and bottom views of the heterostructures, where only the three atomic layers at the interface (h-BN/Te-Bi or In-Se/h-BN) are presented. The energies are listed in Table.~\ref{table4}, which is referenced to the configuration of S1 with FM IMC and downward polarization. Overall, the h-BN substrate has little effect on the IMC in the heterojunction. The layer-projected band structures of h-BN/MnBi$_2$Te$_4$-2L/In$_2$Se$_3$/h-BN heterostructures withdifferent polarizations are shown in Fig.~\ref{S9}(c) and (d). The results demonstrate that the h-BN substrate does not alter the band arrangement between MnBi$_2$Te$_4$-2L and In$_2$Se$_3$, as it is a large bandgap insulator.

~\\

\begin{figure}[htbp]
\centering
\includegraphics[width=0.9\textwidth]{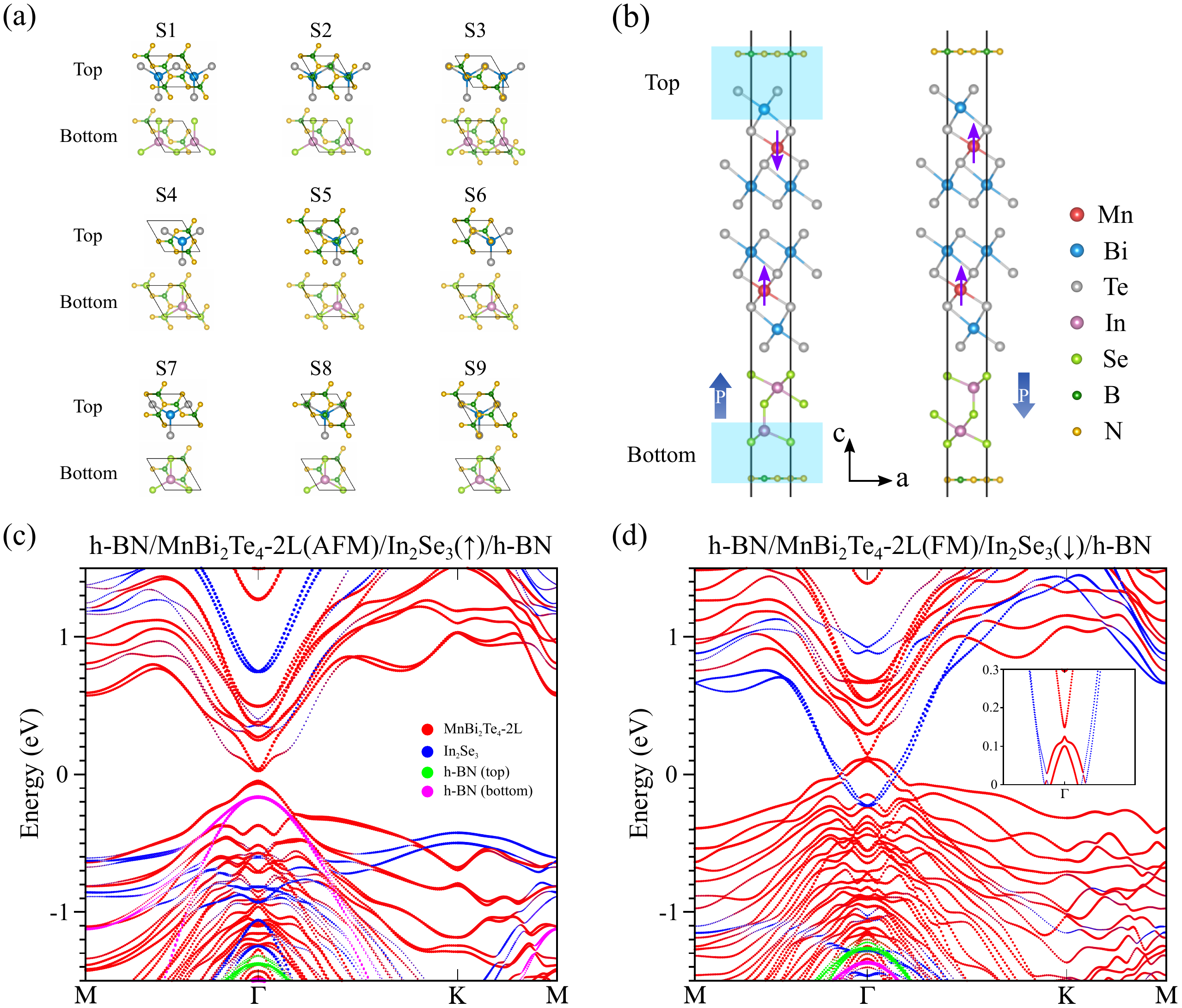}
\caption{Geometric structures and band structure of the sandwich heterojunction of h-BN/MnBi$_2$Te$_4$-based/h-BN. (a) The top view of h-BN/MnBi$_2$Te$_4$-based/h-BN heterostructures with different stacking configuration. Here, only the three atomic layers at the interface are shown. The 'Top' ('Bottom') denote the stacking between the top (bottom) h-BN and MnBi$_2$Te$_4$-2L/In$_2$Se$_3$, respectively. (b) The side view of the lowest energy configuration for h-BN/MnBi$_2$Te$_4$-2L/In$_2$Se$_3$/h-BN heterostructures. Left and right panel shows The side view of the lowest energy configuration for h-BN/MnBi$_2$Te$_4$-2L/In$_2$Se$_3$/h-BN heterostructures with different polarizations. The thin purple arrows denote spins of the Mn ions. While the thick blue arrows denote polarizations of the FE substrate. (c, d) Band structures for the h-BN/MnBi$_2$Te$_4$-2L(AFM)/In$_2$Se$_3(\uparrow)$/h-BN and h-BN/MnBi$_2$Te$_4$-2L(FM)/In$_2$Se$_3(\downarrow)$/h-BN.}
\label{S9}
\end{figure}

\begin{table}
\renewcommand{\arraystretch}{1.16}
\centering
\caption{Effects of substrates on the IMC of MnBi$_2$Te$_4$-2L/In$_2$Se$_3$. $\Delta E$ = $E_{FM}$ $-$ $E_{AFM}$, $E_{FM}$ ($E_{AFM}$) represents the total energy of the FM (AFM) state. Energies (in meV) of different interlayer magnetic states of freestanding h-BN/MnBi$_2$Te$_4$-2L/In$_2$Se$_3$($\downarrow$)/h-BN.}
\label{table4}
\begin{tabular}{p{1.5cm}<{\centering}p{1.2cm}<{\centering}p{1.2cm}<{\centering}p{1.2cm}<{\centering}p{1.2cm}<{\centering}p{1.2cm}<{\centering}p{1.2cm}<{\centering}p{1.2cm}<{\centering}p{1.2cm}<{\centering}}
\toprule[0.7 pt]
\toprule[0.7 pt]
          \multirow {2}*{Staking} & \multicolumn{4}{c}{h-BN/MnBi$_2$Te$_4$-2L/In$_2$Se$_3$($\uparrow$)/h-BN} & \multicolumn{4}{c}{h-BN/MnBi$_2$Te$_4$-2L/In$_2$Se$_3$($\downarrow$)/h-BN} \\
          ~          & FM & AFM &  $\Delta E$ & IMCs & FM & AFM &  $\Delta E$  & IMCs \\
        \hline
        S1 & 65.00  & 64.80 & 0.20 & AFM & 0.00 & 0.17 & -0.17 & FM   \\
        S2 & 70.16  & 69.93 & 0.23 & AFM & 5.04 & 5.19 & -0.15 & FM   \\
        S3 & 75.05  & 74.84 & 0.21 & AFM & 9.69 & 9.84 & -0.15 & FM   \\
        S4 & 68.15  & 67.92 & 0.23 & AFM & 3.54 & 3.71 & -0.17 & FM   \\
        S5 & 76.28  & 76.06 & 0.22 & AFM & 8.95 & 9.12 & -0.17 & FM   \\
        S6 & 78.00  & 77.78 & 0.22 & AFM & 12.37 & 12.53 & -0.16 & FM   \\
        S7 & 73.68  & 73.46 & 0.22 & AFM & 10.13 & 10.29 & -0.16 & FM   \\
        S8 & 78.43  & 78.21 & 0.22 & AFM & 15.15 & 15.30 & -0.15 & FM   \\
        S9 & 83.60  & 83.39 & 0.21 & AFM & 19.20 & 19.37 & -0.17 & FM   \\
        \hline
\toprule[0.7 pt]
\toprule[0.7 pt]
\end{tabular}
\end{table}

\newpage
~\\
~\\
\section*{\Large 10. Stability of different interlayer magnetic couplings in MnBi$_2$Te$_4$-based sandwiches}
\begin{table}[h]
\renewcommand{\arraystretch}{1.16}
\centering
\caption{Energies (in meV) of different interlayer magnetic states for freestanding MnBi$_2$Te$_4$-3L and its sandwich structure In$_2$Se$_3$/MnBi$_2$Te$_4$-3L/In$_2$Se$_3$.
The energies are calculated by taking those of the magnetic ground states as the reference. The state In$_2$Se$_3$($\downarrow$)/MnBi$_2$Te$_4$-3L/In$_2$Se$_3$($\uparrow$)
will not be shown during the ferroeectric transforming as discussed in the main text.}
\label{table5}
\begin{tabular}{p{1.6cm}<{\centering}p{2.4cm}<{\centering}p{2.4cm}<{\centering}p{2.4cm}<{\centering}p{2.4cm}<{\centering}}
\toprule[0.7 pt]
\toprule[0.7 pt]
          \multirow {2}*{IMC} & \multirow {2}*{freestanding} & In$_2$Se$_3$($\uparrow$)/MBT & In$_2$Se$_3$($\uparrow$)/MBT & In$_2$Se$_3$($\downarrow$)/MBT   \\
          ~          &            &      -3L/In$_2$Se$_3$($\uparrow$)    &    -3L/In$_2$Se$_3$($\downarrow$)    &    -3L/In$_2$Se$_3$($\downarrow$)   \\
        \hline
        $\uparrow\uparrow\uparrow$            &    0.47      &      0.17      &     0.00     &    0.17   \\
        $\downarrow\uparrow\uparrow$          &    0.21      &      0.22      &     0.22     &    0.00   \\
        $\uparrow\downarrow\uparrow$          &    0.00      &      0.06      &     0.41     &    0.07   \\
        $\downarrow\downarrow\uparrow$        &    0.21      &      0.00      &     0.20     &    0.23   \\
        $\downarrow\downarrow\downarrow$      &    0.47      &      0.17      &     0.00     &    0.17   \\
        $\uparrow\downarrow\downarrow$        &    0.21      &      0.21      &     0.21     &    0.00   \\
        $\downarrow\uparrow\downarrow$        &    0.00      &      0.06      &     0.42     &    0.08   \\
        $\uparrow\uparrow\downarrow$          &    0.21      &      0.00      &     0.21     &    0.23   \\
        \hline
\toprule[0.7 pt]
\toprule[0.7 pt]
\end{tabular}
\end{table}

\begin{table}[h]
\renewcommand{\arraystretch}{1.16}
\centering
\caption{Energies (in meV) of different interlayer magnetic states of freestanding MnBi$_2$Te$_4$-4L and its sandwich structure In$_2$SSe$_2$/MnBi$_2$Te$_4$-4L/In$_2$Se$_3$.
The energies are calculated by taking those of the magnetic ground states as the reference.}
\label{table6}
\begin{tabular}{p{1.0cm}<{\centering}p{1.8cm}<{\centering}p{2.4cm}<{\centering}p{2.4cm}<{\centering}p{2.4cm}<{\centering}p{2.4cm}<{\centering}}
\toprule[0.7 pt]
\toprule[0.7 pt]
          \multirow {2}*{IMC} & \multirow {2}*{freestanding} & In$_2$SSe$_2$($\uparrow$)/MBT & In$_2$SSe$_2$($\uparrow$)/MBT & In$_2$SSe$_2$($\downarrow$)/MBT  &  In$_2$SSe$_2$($\downarrow$)/MBT \\
          ~          &            &      -4L/In$_2$Se$_3$($\uparrow$)    &    -4L/In$_2$Se$_3$($\downarrow$)    &    -4L/In$_2$Se$_3$($\downarrow$)  &    -4L/In$_2$Se$_3$($\uparrow$) \\
        \hline
        $\uparrow\uparrow\uparrow\uparrow$            &    0.70      &      0.40      &     0.12     &    0.39     &    0.70   \\
        $\downarrow\uparrow\uparrow\uparrow$          &    0.45      &      0.42      &     0.24     &    0.18     &    0.46   \\
        $\uparrow\downarrow\uparrow\uparrow$          &    0.21      &      0.24      &     0.11     &    0.00     &    0.21   \\
        $\downarrow\downarrow\uparrow\uparrow$        &    0.42      &      0.17      &     0.00     &    0.19     &    0.43   \\
        $\uparrow\uparrow\downarrow\uparrow$          &    0.21      &      0.00      &     0.07     &    0.24     &    0.21   \\
        $\downarrow\uparrow\downarrow\uparrow$        &    0.00      &      0.07      &     0.22     &    0.05     &    0.00   \\
        $\uparrow\downarrow\downarrow\uparrow$        &    0.22      &      0.21      &     0.35     &    0.20     &    0.21   \\
        $\downarrow\downarrow\downarrow\uparrow$      &    0.45      &      0.17      &     0.21     &    0.44     &    0.45   \\
        $\downarrow\downarrow\downarrow\downarrow$    &    0.70      &      0.40      &     0.12     &    0.38     &    0.70   \\
        $\uparrow\downarrow\downarrow\downarrow$      &    0.45      &      0.43      &     0.24     &    0.16     &    0.45   \\
        $\downarrow\uparrow\downarrow\downarrow$      &    0.21      &      0.24      &     0.11     &    0.00     &    0.21   \\
        $\uparrow\uparrow\downarrow\downarrow$        &    0.42      &      0.17      &     0.00     &    0.18     &    0.43   \\
        $\downarrow\downarrow\uparrow\downarrow$      &    0.21      &      0.00      &     0.09     &    0.25     &    0.21   \\
        $\uparrow\downarrow\uparrow\downarrow$        &    0.00      &      0.05      &     0.21     &    0.05     &    0.00   \\
        $\downarrow\uparrow\uparrow\downarrow$        &    0.22      &      0.22      &     0.31     &    0.21     &    0.21   \\
        $\uparrow\uparrow\uparrow\downarrow$          &    0.45      &      0.16      &     0.21     &    0.45     &    0.45   \\
        \hline
\toprule[0.7 pt]
\toprule[0.7 pt]
\end{tabular}
\end{table}

\newpage

~\\
~\\
\section*{\Large 11. Kinetics pathways of ferroelectric  transforming for MnBi$_2$Te$_4$-based heterojunctions}

\begin{figure}[htbp]
\centering
\includegraphics[width=0.95\textwidth]{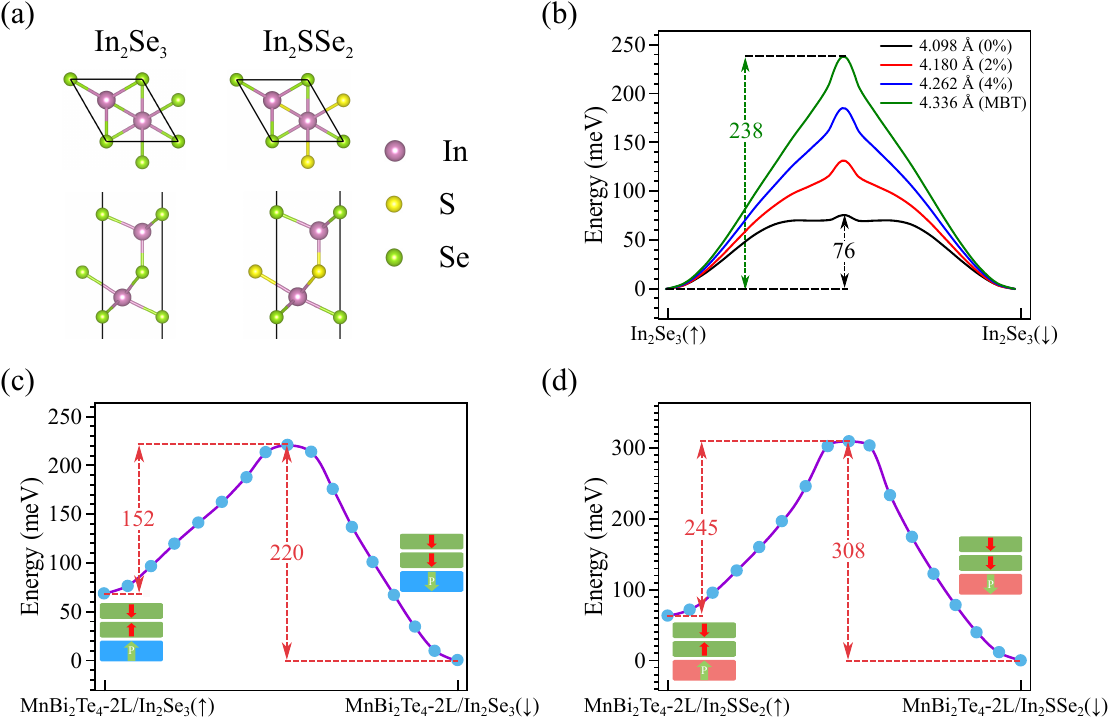}
\caption{(a) Geometric structures of In$_2$Se$_3$ and In$_2$SSe$_2$ monolayer. The top and bottom panels show the top and side views, respectively. (b) Energy barriers for the freestanding In$_2$Se$_3$ monolayer in different lattice constants. (c, d) The kinetic pathways of ferroelectric transforming for MnBi$_2$Te$_4$-2L/In$_2$Se$_3$ and MnBi$_2$Te$_4$-2L/In$_2$SSe$_2$, respectively. Energy barriers are given in meV.}
\label{S11}
\end{figure}

As discussed in the main text,  sandwiching the MnA$_2$Te$_4$ thin films in between two different ferroelectric layers is a way of realizing the quadruple magnetic states. Here, we choose the ferroelectric monolayers of In$_2$Se$_3$ and In$_2$SSe$_2$ as substrates to tune the IMC of MAT films. Geometric structures of In$_2$Se$_3$ and In$_2$SSe$_2$ monolayers are shown in Fig.~\ref{S11}(a). The latter can be obtained by replacing the central Se atom of the former by a S atom. Figures~\ref{S11} (c) and (d) show the kinetic pathways of ferroelectric transforming for MnBi$_2$Te$_4$-2L/In$_2$Se$_3$ and MnBi$_2$Te$_4$-2L/In$_2$SSe$_2$. Both systems has an asymmetric energy barriers due to the interface effect. For the MnBi$_2$Te$_4$-2L/In$_2$Se$_3$ (MnBi$_2$Te$_4$-2L/In$_2$SSe$_2$) system, the pathways of P($\uparrow$)-to-P($\downarrow$) and P($\downarrow$)-to-P($\uparrow$) has a barrier height are about 152 (245) and 220 (308) meV, respectively. The barrier heights of the MnBi$_2$Te$_4$-2L/In$_2$SSe$_2$ system are higher than those of the MnBi$_2$Te$_4$-2L/In$_2$Se$_2$. This trend is important in realizing the quadruplet magnetic state (see Fig.3 (c) in the main text).

\begin{figure}[htbp]
\centering
\includegraphics[width=0.95\textwidth]{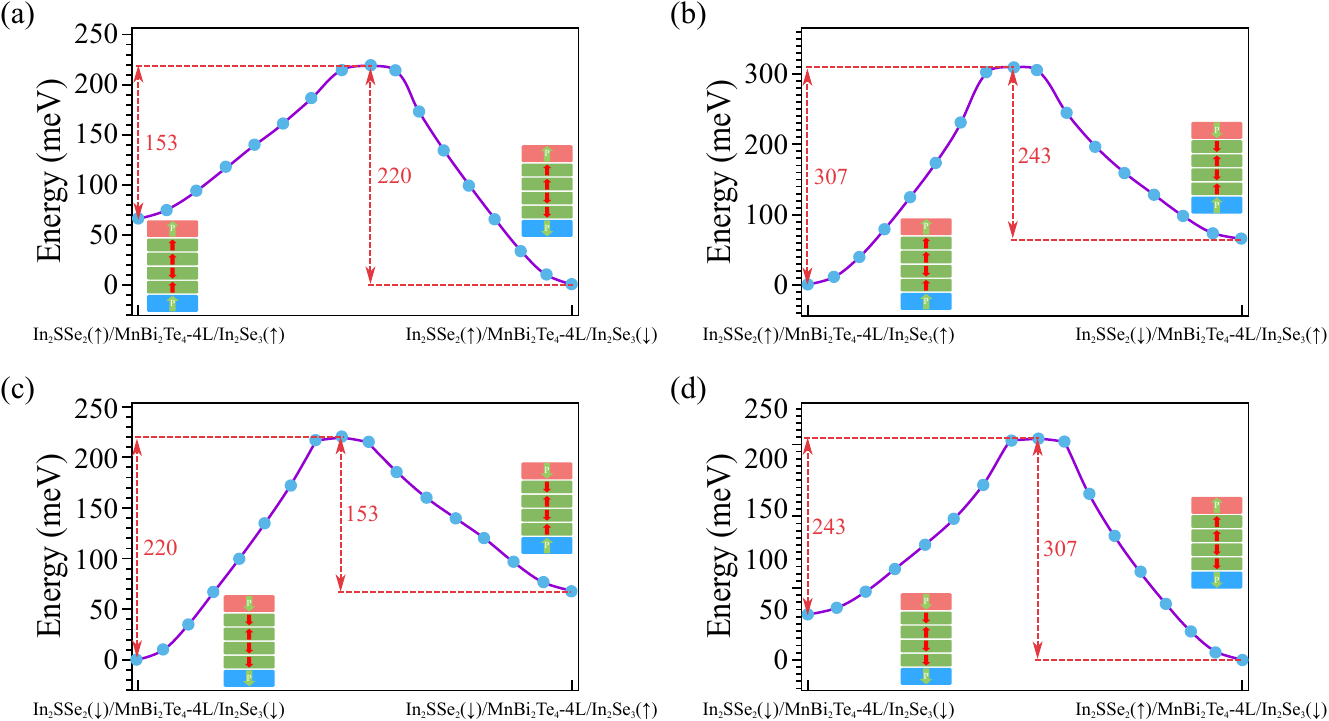}
\caption{Comparison of barrier heights (in meV) for flipping polarizations in different layers in In$_2$SSe$_2$/MnBi$_2$Te$_4$-4L/In$_2$Se$_3$.
}
\label{S12}
\end{figure}

Furthermore, we investigate the kinetic pathways for the transformings between the quadruple states in the In$_2$SSe$_2$/MnBi$_2$Te$_4$-4L/In$_2$Se$_3$ system by
examining the barriers of all possible pathways. Fig.~\ref{S12} shows a comparison of the energy barriers with respect to flip the polarization of In$_2$Se$_3$ and
In$_2$SSe$_2$ layers. For the state with upward polarization for both the top and bottom ferroelectric layers, i.e., InSSe$_2$($\uparrow$)/MnBi$_2$Te$_4$-4L/In$_2$Se$_3$($\uparrow$),
flipping the polarizations of InSSe$_2$ layer has a barrier height that is about 154 meV higher than that the In$_2$Se$_3$ layer. Similarly, for
InSSe$_2$($\downarrow$)/MnBi$_2$Te$_4$-4L/In$_2$Se$_3$($\downarrow$) state, flipping the polarizations of InSSe$_2$ layer has a barrier height that is about
23 meV higher than that of the In$_2$Se$_3$ layer. Therefore,  the polarizations of In$_2$Se$_3$ have priority to those of the In$_2$SSe$_2$ to flip under external electric fields, which are desired for realizing quadruple magnetic states.

\newpage
\section*{\Large 12. Topological properties of MnBi$_2$Te$_4$ thin films with different interlayer magnetic states}

\begin{figure}[htbp]
\centering
\includegraphics[width=0.95\textwidth]{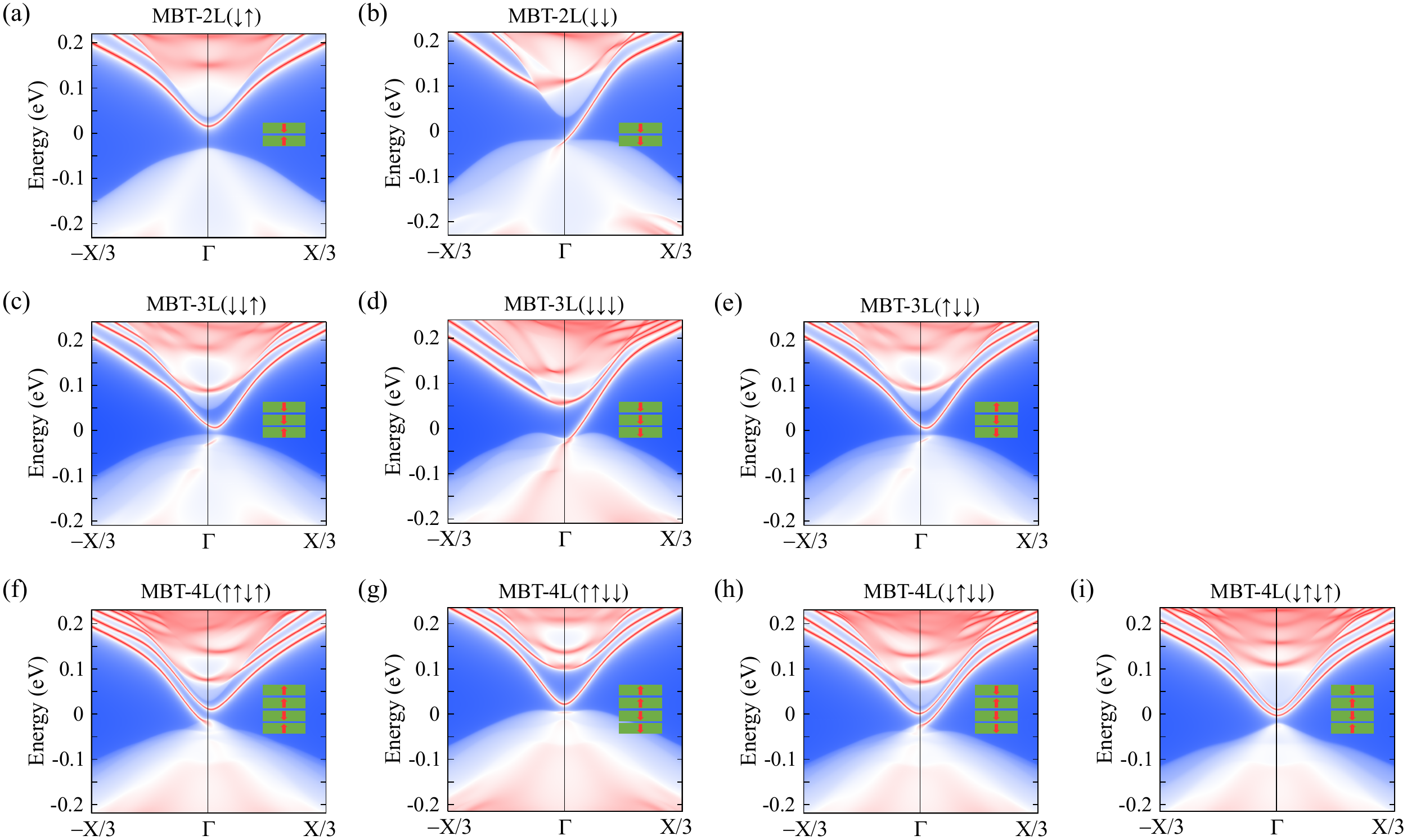}
\caption{Edge states of freestanding MnBi$_2$Te$_4$ thin films.
The red arrows in the inset represent the magnetizations of the Mn atoms.}
\label{S13}
\end{figure}

The topological properties of the MnBi$_2$Te$_4$ system are strongly related to the IMCs. Fig.~\ref{S13} shows the IMCs dependence of edge states for freestanding MnBi$_2$Te$_4$ thin films. For the bilayer systems, we have gapped (gapless) edge states and 0 (1) for
the Chen number ($C$) for the interlayer AFM (FM) ordering (see Figs.~\ref{S13} a and b). For the trilayers,  we obtain $C$ = 0, 1,
and 0 for $T_1$ ($\downarrow\downarrow\uparrow$), $T_2$ ($\downarrow\downarrow\downarrow$), and $T_3$ ($\uparrow\downarrow\downarrow$), respectively.
As shown in Fig.~\ref{S13} (c)-(e), the MnBi$_2$Te$_4$-3L in magnetic state $T_2$ ($\downarrow\downarrow\downarrow$) has gapless edge states
in the bulk band gap, which are not seen for $T_1$($\downarrow\downarrow\uparrow$) and $T_3$($\uparrow\downarrow\downarrow$). For MnBi$_2$Te$_4$ quad-layers,  we obtain $C$ = 1, 0, -1, and 0 for $Q_1$ ($\uparrow\uparrow\downarrow\uparrow$), $Q_2$ ($\uparrow\uparrow\downarrow\downarrow$),
$Q_3$ ($\downarrow\uparrow\downarrow\downarrow$), and $Q_4$ ($\downarrow\uparrow\downarrow\uparrow$), respectively.
Correspondingly, $Q_1$ ($\uparrow\uparrow\downarrow\uparrow$) and $Q_3$ ($\downarrow\uparrow\downarrow\downarrow$) states have gapless edge states
with opposite chiralities.

\begin{figure}[htbp]
\centering
\includegraphics[width=0.95\textwidth]{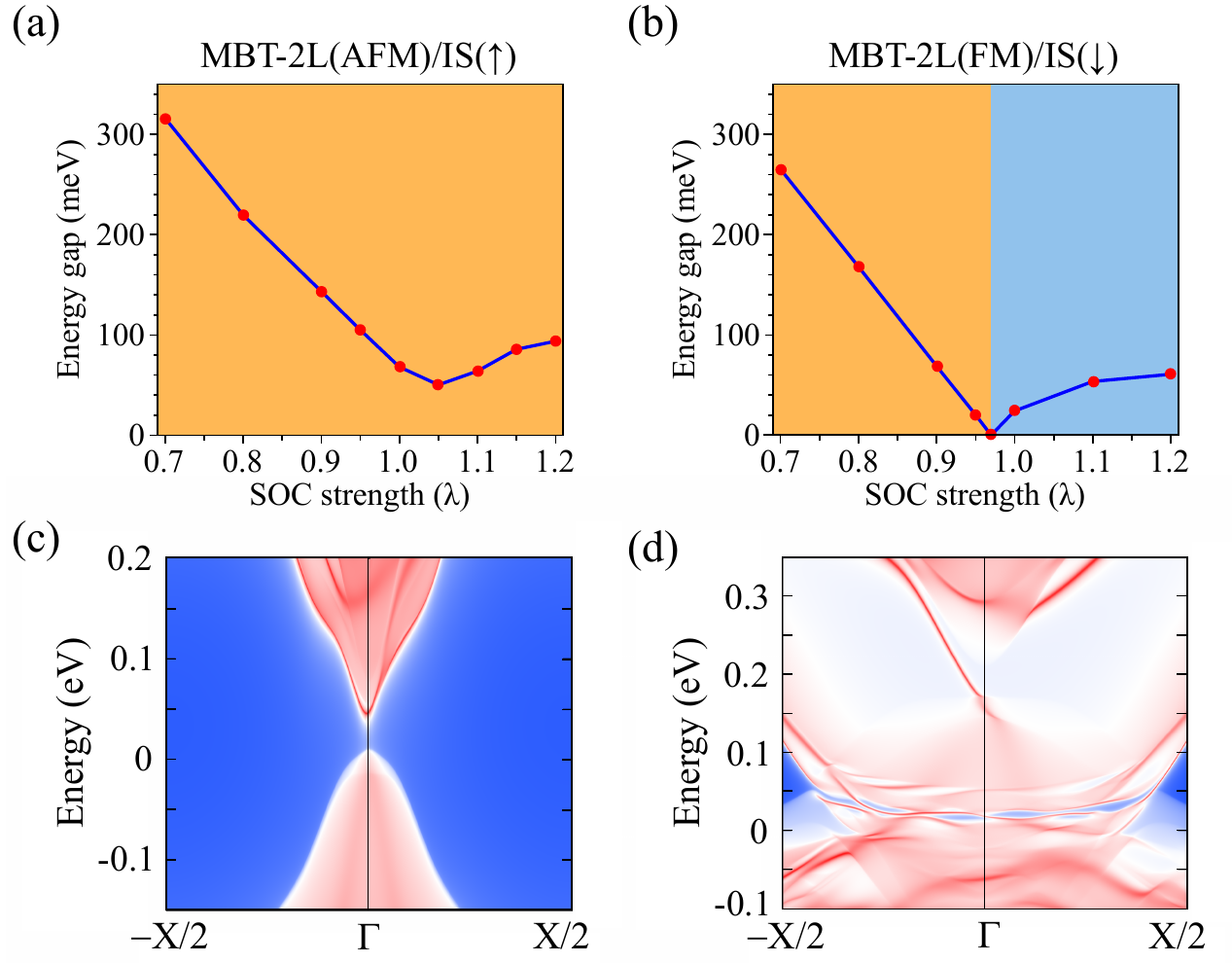}
\caption{Topological properties of MnBi$_2$Te$_4$-2L/In$_2$Se$_3$. (a, b) Band gap evolution for MnBi$_2$Te$_4$-2L on the In$_2$Se$_3$ substrate
  (P$\uparrow$ and P$\downarrow$) with respect to SOC strength ($\lambda$) from 0.7 to 1.2. (c, d) Edge states  of MnBi$_2$Te$_4$-2L/IS($\uparrow$)
  and MnBi$_2$Te$_4$-2L/IS($\downarrow$), respectively. }
\label{S14}
\end{figure}

We expect that the ferroelectrically tunable Chern numbers can be seen in MnBi$_2$Te$_4$-based heterostructure. Fig.~\ref{S14} shows
the topological properties of MnBi$_2$Te$_4$-2L/In$_2$Se$_3$. For the bilayer systems, the topological properties of MnBi$_2$Te$_4$ remain unchanged upon
interfacing, i.e., $C = 1$ for the FM state and $C = 0$ for the AFM state. The evolution of the band gap with respect to the SOC strength ($\lambda$) is
shown in Figs.~\ref{S14}(a) and (b). For the MnBi$_2$Te$_4$-2L(AFM)/In$_2$Se$_3$($\uparrow$) state, the gap decreases with increasing $\lambda$ and
reaches a minimum (51 meV) at $\lambda$ = 1.05. Then it increases as the SOC increases. However, for the MnBi$_2$Te$_4$-2L(FM)/In$_2$Se$_3$($\downarrow$)
state, the gap decreases and eventually closes at $\lambda$ = 0.97 and then opens a gap with increasing $\lambda$, which shows a band inversion and
topological phase transition. Furthermore,  edge state calculations show that  MnBi$_2$Te$_4$-2L(AFM)/In$_2$Se$_3$($\uparrow$) state has no gapless states (Fig.~\ref{S14}c). Whereas MnBi$_2$Te$_4$-2L(FM)/In$_2$Se$_3$($\downarrow$) state has a gapless edge state in the bulk energy gap (see Fig.~\ref{S14}d).

\end{document}